\documentclass[12pt]{article}
\usepackage[english,activeacute]{babel}
\usepackage{natbib}
\usepackage{comment}
\usepackage{float}
\usepackage[hidelinks]{hyperref}
\usepackage{mathrsfs}
\usepackage{enumitem}
\usepackage[font={small,it}]{caption}
\usepackage{amsmath,amsfonts,amsthm,amssymb}
\usepackage{bm,rotating,multirow,dsfont,graphicx}
\usepackage[usenames, dvipsnames]{color}
\usepackage{url}
\usepackage{multicol}
\usepackage{multirow}
\usepackage[T1]{fontenc}
\usepackage{flafter}
\usepackage{appendix}
\usepackage{subfigure}
\usepackage{xcolor}
\usepackage{soul}
\usepackage{setspace}
\usepackage{booktabs}
\usepackage{algorithm}
\usepackage{algpseudocode}
\usepackage{mathtools}
\makeatletter
\def\hlinewd#1{%
	\noalign{\ifnum0=`}\fi\hrule \@height #1 %
	\futurelet\reserved@a\@xhline}
\makeatother
\def\spacingset#1{\renewcommand{\baselinestretch}{#1}\small\normalsize}\spacingset{1}
\def\@roman#1{\romannumeral #1}

\begin{document}

\title{Spherical latent space models \\ for social network analysis}

\date{}

\author{
    Juan Sosa\footnote{Corresponding author: jcsosam@unal.edu.co.}\qquad \qquad 
    Carlos Nosa\footnote{Corresponding author: cnosa@unal.edu.co.} \\
    Universidad Nacional de Colombia, Colombia
}

\maketitle


\begin{abstract}
This article introduces a spherical latent space model for social network analysis, embedding actors on a hypersphere rather than in Euclidean space as in standard latent space models. The spherical geometry facilitates the representation of transitive relationships and community structure, naturally captures cyclical patterns, and ensures bounded distances, thereby mitigating degeneracy issues common in traditional approaches. Bayesian inference is performed via Markov chain Monte Carlo methods to estimate both latent positions and other model parameters. The approach is demonstrated using two benchmark social network datasets, yielding improved model fit and interpretability relative to conventional latent space models.
\end{abstract}


\spacingset{1.1} 

\newpage

\section{Introduction}

The study of complex systems through their interconnected elements has become central to numerous scientific disciplines. Such systems are commonly represented as networks or graphs, comprising autonomous actors (nodes) and the relational ties (edges) between them. These structures reveal patterns underlying phenomena ranging from disease transmission to social collaboration. The breadth of applications underscores the versatility of network models, which can represent entities as diverse as individuals, organizations, or nations, linked by relationships such as friendships, collaborations, or information exchange, respectively.

Modern statistical network analysis has three primary objectives: (1) identifying and summarizing structural patterns, (2) developing stochastic models to explain network formation processes, and (3) predicting unobserved or future connections based on network properties and actor attributes (see \citealt{posfai2016network}, \citealt{al2017python}, \citealt{newman2018networks}, \citealt{brugere2018network}, and \citealt{drobyshevskiy2019random} for comprehensive reviews of network definitions, properties, methods, models, and processes across multiple fields). A central challenge lies in formulating statistical models (rather than deterministic mechanisms) that capture the inherent dependencies among ties, particularly reciprocity and transitivity. Traditional approaches, such as the exponential random graph models (ERGMs) of \cite{robins_introduction_2007}, address these dependencies but are often susceptible to degeneracy and limited interpretability.

In statistical network modeling, latent space models offer a compelling alternative (see \citealt{goldenberg2010survey}, \citealt{salter2012review}, \citealt{matias2014modeling}, \citealt{sosa_review_2021}, and \citealt{kaur2023latent} for comprehensive reviews of latent space modeling and related statistical approaches). In particular, the latent space model, introduced in the foundational work of \cite{hoff_latent_2002}, embeds actors in an unobserved, typically low-dimensional, latent space (often referred to as the social space) where the probability of a tie is modeled as a function of the distance or similarity measure between actors in that space. This formulation naturally captures transitivity, as actors positioned closer together in latent space are more likely to form connections. The geometric structure of these models renders them both computationally tractable and interpretable. Nonetheless, they may be inadequate for networks generated by highly nonlinear processes or exhibiting complex geometric features beyond the representational capacity of Euclidean spaces.

Latent space models have been extensively extended to a wide range of settings. Beyond simple distance-based formulations, alternative latent structures have been proposed, including projection models (\citealt{hoff_latent_2002}), bilinear models (\citealt{hoff_bilinear_2005}, \citealt{hoff2021additive}), and eigenmodels (\citealt{hoff2008}, \citealt{hoff2009multiplicative}, \citealt{minhas2019inferential}). These frameworks have been generalized to accommodate multiple structural features, such as multilayer networks (\citealt{hoff2011hierarchical}, \citealt{kivela_multilayer_2014}, \citealt{gollini2016joint}, \citealt{salter2017latent}, \citealt{dangelo2019latent}, \citealt{zhang2020flexible}, \citealt{sosa2021latent}, \citealt{sosa_betancourt_multilayer_2022}, \citealt{d2023model}) and dynamic network settings (\citealt{durante2014nonparametric}, \citealt{sewell_dynamic_2015}, \citealt{hoff2015multilinear}, \citealt{sewell2016latent}, \citealt{sewell2017latent}, \citealt{kim2018review}, \citealt{he2019multiplicative}, \citealt{guhaniyogi2020joint}). They have also been adapted to perform diverse statistical tasks, including covariate testing (\citealt{austin2013covariate}, \citealt{fosdick2015testing}, \citealt{wang2023joint}), clustering (\citealt{nowicki_stochastic_2001}, \citealt{krivitsky2009representing}, \citealt{ryan2017bayesian}, \citealt{rastelli2018optimal}, \citealt{ng2022model}), modeling bipartite relational data (\citealt{friel2016interlocking}), deduplication and record linkage (\citealt{sosa_deduplication_relational_2022}, \citealt{sosa2023bayesian}), and the study of influence and diffusion processes (\citealt{xu2018alternative}, \citealt{sanchez_gutierrez_influence_2025}). Moreover, significant efforts have focused on developing faster computational alternatives to improve scalability and efficiency (\citealt{hunter2012computational}, \citealt{raftery2012fast}, \citealt{salter2013variational}, \citealt{caimo2016bayesian}, \citealt{ma2020universal}, \citealt{aliverti2022stratified}, \citealt{spencer2022faster}, \citealt{rastelli2024computationally}).

One way to address the limitations of fitting latent space models in Euclidean space is to embed actors on a hypersphere rather than in a flat geometry. This idea is not entirely new; it has been recently explored in political science to model phenomena that standard voting space approaches cannot capture when analyzing roll call data \citep{yu2020spherical}. To the best of our knowledge, however, spherical embeddings have not yet been applied to latent space models for general network analysis. The spherical geometry offers several advantages: it defines a bounded space that prevents extreme discrepancies in tie probabilities, naturally represents directional clustering, and provides additional modeling flexibility through curvature and cyclic structures. We develop a Bayesian inference framework for estimating latent positions and model parameters using Markov chain Monte Carlo (MCMC; e.g., \citealt{gamerman2006markov}) methods. Finally, we demonstrate our approach on benchmark real-world network datasets as in \cite{hoff_latent_2002}, showing improved model fit and interpretability over the traditional Euclidean formulation.

The remainder of this article is organized as follows. Section 2 outlines the fundamentals of network structures and their characterization. Section 3 reviews related work on latent space modeling and introduces the proposed spherical latent space framework along with its statistical properties. Section 4 details the Bayesian inference procedure and computational strategies. Section 5 presents the experimental results, and Section 6 concludes with a discussion of implications and avenues for future research.

\section{Fundamentals}

Network data consist of actors, nodal attributes measured on individuals, and dyadic attributes measured on actor pairs. Dyadic variables may be binary, indicating the presence or absence of a tie (binary networks), or weighted, quantifying relational aspects (valued networks). Relations can be undirected, with a single value per pair, or directed, with two values capturing each actor’s perspective. Networks are typically represented as graphs or adjacency matrices $\mathbf{Y} = [y_{i,j}]$, where $y_{i,j}$ denotes the presence, absence, or weight of a tie. Self-loops generate structural zeros on the diagonal, and the symmetry or asymmetry of $\mathbf{Y}$ indicates whether the network is undirected or directed.

Statistical network analysis methods fall into three categories: descriptive, which visualize and quantify structure; modeling and inference, which explain network formation; and process, which study how interactions influence attributes. This work focuses on modeling and inference, emphasizing structural features such as cohesion, connectivity, and assortativity that reflect dependencies like reciprocity and clustering. Accounting for these dependencies is crucial for constructing realistic statistical models. For comprehensive reviews of network properties and measures, see \cite{menczer2020first} and \cite{kolaczyk_statistical_2020}.

\subsection{Vertex characteristics}

Characterizing a network often begins with describing its vertices. The degree of a vertex, the number of incident edges, identifies highly connected nodes. In most real-world networks, degree distributions are strongly right-skewed and often follow a power-law, unlike those of random graphs. Examining the relationship between a vertex’s degree and the average degree of its neighbors reveals how nodes of different connectivity link to each other. Furthermore, centrality measures assess the relative importance of vertices. For instance, closeness centrality favors vertices near many others, while betweenness centrality highlights those lying on numerous shortest paths. Most centrality metrics rely on geodesic distance, though many variations exist.

\subsection{Network structure}

Two key structural properties of a network are cohesion and connectivity. Cohesion can be assessed by checking whether the network is connected (every vertex is reachable from every other) or complete (every vertex is linked to every other), and by enumerating specific subgraphs such as dyads, triads, or cliques. Quantitative measures include the density, the proportion of observed to potential edges, which reflects proximity to completeness, and the clustering coefficient, the proportion of connected triples that form triangles, which captures transitivity. These measures can also be computed locally for individual vertices. Connectivity is further examined through resilience, which evaluates how vertex removal affects path existence.

Networks often exhibit homophily or assortative mixing, where vertices preferentially connect to others with similar attributes. Homophily helps explain patterns such as transitivity, balance, and cohesive subgroups. Assortativity coefficients, variations of correlation coefficients, quantify its extent and often summarize degree correlations between adjacent vertices. An extreme manifestation of homophily is community structure, where groups of vertices have dense internal and sparse external connections. Communities can be detected without external information using methods such as hierarchical clustering, which groups vertices based on similarity, and spectral partitioning, which iteratively applies eigen decomposition of the graph Laplacian.

\subsection{Statistical models for networks}

Generally speaking, a statistical network model is a probability distribution on a sociomatrix $\mathbf{Y}$ indexed by an unknown parameter $\boldsymbol{\theta} \in \Theta$, $p(\mathbf{Y} \mid \boldsymbol{\theta})$. Rather than merely visualizing and describing the topological characteristics of a network, statistical models seek to capture the essential aspects of the stochastic mechanism by which the network may have arisen. They enable testing the significance of predefined structural features, assessing associations between node or edge attributes and the overall network structure, and imputing missing observations. Unlike deterministic or purely algorithmic models, statistical approaches also quantify the uncertainty associated with unknown quantities. Importantly, the very nature of network data induces dependencies both between actors and between ties. Accounting for these dependencies is indispensable for formulating statistically sound and substantively meaningful network models.

\subsection{Latent space models}

Latent space models are a widely used approach to network modeling, originally introduced by \citet{hoff_latent_2002} and further developed by \citet{handcock2007modeling} and \citet{krivitsky2009representing}. In this framework, each node is assigned a latent position in a Euclidean space, and the probability of an edge between two nodes depends on their proximity in that space. These models can be viewed as generalized linear models with random effects, allowing them to capture complex network structures. For conditionally independent $y_{i,j}$ in the undirected binary case, the interaction probabilities are  
\[
\textsf{Pr}(y_{i,j} = 1 \mid \mathbf{x}_{i,j}, \boldsymbol{\beta}, \zeta_{i,j}) = \textsf{logit}^{-1}(\mathbf{x}^\top_{i,j}\boldsymbol{\beta} + \zeta_{i,j}), \qquad 1 \leq i < j \leq n,
\]  
where $\boldsymbol{\beta}$ are fixed effects, $\mathbf{x}^\top_{i,j}\boldsymbol{\beta}$ represents the contribution of covariates $\mathbf{x}_{i,j}$, and $\zeta_{i,j}$ captures unobserved effects, and $\textsf{logit}(x) = \log\frac{x}{1-x}$ is the link function. As noted in \citet{hoff2008}, following foundational results from \cite{hoover1982} and \cite{aldous1985}, a jointly exchangeable random effects matrix $[\zeta_{i,j}]$ can be expressed as $\zeta_{i,j} = h(\mathbf{z}_i, \mathbf{z}_j)$, where $h(\cdot,\cdot)$ is a symmetric function and $\mathbf{z}_1, \ldots, \mathbf{z}_n$ are independent latent variables (vectors). The function $h(\cdot,\cdot)$ is central to modeling relational data, and various formulations have been proposed. In this paper, we focus on the distance formulation for undirected binary networks. For a comprehensive review, see \citet{sosa_review_2021}.

\section{Spherical latent space models}\label{section3}

Consider modeling relational data from an undirected binary network represented by an $n \times n$ adjacency matrix $\mathbf{Y} = [y_{i,j}]$, where each entry $y_{i,j} \in \{0,1\}$ indicates the presence or absence of a relation between actors $i$ and $j$. A widely used approach in this context is the standard latent distance model \citep{hoff_latent_2002}, which assumes that each actor $i$ occupies an unobserved position $\mathbf{z}_i \in \mathbb{R}^d$, often referred to as the latent or social space. Conditional on these positions, ties are modeled as  
\[
y_{i,j} \mid \eta_{i,j} \overset{\text{ind}}{\sim} \textsf{Ber}\big(\textsf{logit}^{-1}(\eta_{i,j})\big), \qquad
\eta_{i,j} = \alpha - \textsf{d}(\mathbf{z}_i, \mathbf{z}_j),
\]  
where $\mathbf{z}_i$ and $\mathbf{z}_j$ are latent position vectors capturing unobserved social characteristics, $\textsf{d}(\cdot,\cdot)$ is a distance function, typically $\textsf{d}(\mathbf{z}_i, \mathbf{z}_j) = \| \mathbf{z}_i - \mathbf{z}_j \|$ with $\|\cdot\|$ denoting the Euclidean norm in $\mathbb{R}^d$, and $\alpha$ is an intercept representing the baseline log odds of a tie when two actors occupy the same latent position. The subtraction of $\textsf{d}(\mathbf{z}_i, \mathbf{z}_j)$ from $\alpha$ reflects the assumption that the probability of a tie decreases as the distance between latent positions increases, encouraging connections between actors located close together in latent space. This property naturally induces transitivity and clustering, as actors close to a common neighbor are also likely to be close to each other.

Alternatively, for modeling in a spherical latent space, the probability of a tie can be parameterized using the inner product between latent vectors, capturing proximity through angular similarity:
\[
\eta_{i,j} = \textsf{logit} \, \textsf{P}(y_{i,j} = 1 \mid \alpha,\beta, \mathbf{z}_i, \mathbf{z}_j) 
= \alpha + \beta \, \langle \mathbf{z}_i, \mathbf{z}_j \rangle,
\]
where $\mathbf{z}_i, \mathbf{z}_j \in \mathbb{S}^{d-1}$ are unit vectors on the $(d-1)$-dimensional sphere, $\langle \mathbf{z}_i, \mathbf{z}_j \rangle = \mathbf{z}_i^\top \mathbf{z}_j$ denotes the standard inner product in $\mathbb{S}^{d-1}$, $\alpha$ is an intercept representing the baseline log odds of a tie when the latent vectors are orthogonal, and $\beta$ is a scaling parameter controlling how strongly angular similarity influences tie formation. Geometrically, $\beta$ determines how sharply the log odds change as the angle between $\mathbf{z}_i$ and $\mathbf{z}_j$ decreases: larger values of $\beta$ make the probability of a tie increase more rapidly as the two latent positions align. Lastly, we emphasize that, analogous to the Euclidean latent distance model, the latent positions $\mathbf{z}_1,\ldots,\mathbf{z}_n$ serve as embeddings of actors that capture unobserved affinities, providing a foundation for visualization, clustering, and structural analysis of the network.

We constrain the latent positions to the $(d-1)$-dimensional unit sphere $\mathbb{S}^{d-1}$, which fixes the origin and scale of the configuration, removing translation and scaling invariances, reducing each position’s degrees of freedom from $d$ to $d-1$, and thus ensuring identifiability up to a global rotation (see Section 3.1 dor details). On $\mathbb{S}^{d-1}$, the inner product $\langle \mathbf{z}_i, \mathbf{z}_j \rangle = \cos(\gamma_{i,j})$ relates directly to the geodesic distance $\gamma_{i,j}$. We use the cosine similarity $\langle \mathbf{z}_i, \mathbf{z}_j \rangle$ rather than $\gamma_{i,j} = \arccos\,\langle \mathbf{z}_i, \mathbf{z}_j \rangle$ to avoid the numerical instability of the $\arccos$ gradient. The spherical formulation yields a compact latent space, preventing positions from drifting to arbitrarily large norms as in Euclidean spaces, thereby avoiding degeneracy and improving numerical stability.

In \citet[Sec.~2.2]{hoff_latent_2002}, the projection model, which is similar in spirit to our approach, specifies that
$$
\eta_{i,j} = \alpha + \mathbf{z}_i^\top \frac{\mathbf{z}_j}{\|\mathbf{z}_j\|},
$$
where similarity is given by the inner product between $\mathbf{z}_i$ and the normalized $\mathbf{z}_j$. This formulation operates in an unconstrained Euclidean space, applies normalization asymmetrically, and is subject to translation and scaling indeterminacies. In contrast, our spherical formulation constrains all latent positions to $\mathbb{S}^{d-1}$, with $\|\mathbf{z}_i\| = 1$ for all $i$, so that $\langle \mathbf{z}_i, \mathbf{z}_j \rangle = \cos(\gamma_{ij})$, where $\gamma_{ij}$ is the geodesic distance. As stated above, such symmetry removes scale and translation non-identifiabilities, reduces each position to $d-1$ free parameters, ensures identifiability up to a global rotation, preserves numerical stability, and prevents norms from diverging.

Given the conditional independence of connections given the intercept term and latent positions, the likelihood is  
$$
\mathcal{L}(\alpha, \beta, \mathbf{Z}) = \prod_{i < j} p_{i,j}^{\,y_{i,j}} \left(1 - p_{i,j}\right)^{1 - y_{i,j}},
$$
where $p_{i,j} = \textsf{logit}^{-1}\left(\alpha + \beta\,\langle \mathbf{z}_i, \mathbf{z}_j \rangle \right)$ is the interaction probability between nodes $i$ and $j$, and $\mathbf{Z} \in \mathbb{S}^{(d-1) \times n}$ is the matrix of latent positions with $\|\mathbf{z}_i\|=1$. Recall that the inner product $\langle \mathbf{z}_i, \mathbf{z}_j \rangle$ equals $\cos(\gamma_{i,j})$, where $\gamma_{i,j}$ is the geodesic distance on the sphere, so similarity is modeled directly via cosine values rather than angular distances. Thus, the log-likelihood takes the form  
$$
\ell(\alpha,\beta,\mathbf{Z}) 
= \sum_{i < j} \left[ y_{i,j} \,\eta_{i,j} - \log \left( 1 + e^{\eta_{i,j}} \right) \right],
$$
with $\eta_{i,j} = \alpha + \beta\,\langle \mathbf{z}_i, \mathbf{z}_j \rangle$, for which the corresponding gradients are given by  
\begin{align*}
    \frac{\partial}{\partial \mathbf{z}_k} \,\ell(\alpha, \beta, \mathbf{Z}) 
    &= \sum_{i \neq k} \left[ y_{i,k} - \textsf{expit}(\eta_{i,k}) \right] 
       \frac{\partial \eta_{i,k}}{\partial \mathbf{z}_k}, \\
    \frac{\partial}{\partial \alpha} \,\ell(\alpha, \beta, \mathbf{Z}) 
    &= \sum_{i < j} \left[ y_{i,j} - \textsf{expit}(\eta_{i,j}) \right], \\
    \frac{\partial}{\partial \beta} \,\ell(\alpha, \beta, \mathbf{Z}) 
    &= \sum_{i < j} \left[ y_{i,j} - \textsf{expit}(\eta_{i,j}) \right] 
       \langle \mathbf{z}_i, \mathbf{z}_j \rangle.
\end{align*}
Analogous expressions can be derived for the Euclidean latent distance model.

\subsection{Model properties}

Latent space models are non-identifiable because the linear predictor is invariant under specific transformations of the latent space. In particular, the likelihood function $\mathcal{L}$ remains unchanged under certain transformations applied to the $d \times n$ matrix $\mathbf{Z}$ containing the latent positions. Such transformations generate equivalence classes of configurations of $\mathbf{Z}$ that produce identical likelihood values.

In the Euclidean formulation, the model is invariant under isometries in $\mathbb{R}^d$ of the form $\mathbf{Z} \mapsto \mathbf{Z}' = \mathbf{Q} \mathbf{Z} + \mathbf{b}\mathbf{1}_n^\top$, where $\mathbf{Q} \in \mathbb{R}^{d\times d}$ is orthogonal, representing a rotation or reflection, $\mathbf{b} \in \mathbb{R}^d$ is a translation vector applied uniformly to all points, and $\mathbf{1}_n \in \mathbb{R}^n$ is a column vector of ones. Indeed, for transformed latent positions $\mathbf{z}_i' = \mathbf{Q} \mathbf{z}_i + \mathbf{b}$ and $\mathbf{z}_j' = \mathbf{Q} \mathbf{z}_j + \mathbf{b}$, we have
\[
\|\mathbf{z}_i' - \mathbf{z}_j'\| 
= \| (\mathbf{Q} \mathbf{z}_i + \mathbf{b}) - (\mathbf{Q} \mathbf{z}_j + \mathbf{b}) \|
= \|\mathbf{Q} (\mathbf{z}_i - \mathbf{z}_j)\| 
= \|\mathbf{z}_i - \mathbf{z}_j\|,
\]
since $\|\mathbf{Q} \mathbf{z}\| = \|\mathbf{z}\|$, for any $\mathbf{z} \in \mathbb{R}^d$ when $\mathbf{Q}$ is orthogonal. Hence, all pairwise distances are preserved, and therefore, $\mathcal{L}(\alpha, \mathbf{Z}) = \mathcal{L}(\alpha, \mathbf{Z}')$, inducing equivalence classes of $\mathbf{Z}$ that yield identical likelihoods and making the latent positions non-identifiable without further constraints.

In the spherical formulation, the latent positions are constrained to the $(d-1)$-dimensional unit sphere $\mathbb{S}^{d-1} \subset \mathbb{R}^d$, fixing both the origin and the scale of the configuration, removing translation and scaling invariances, and reducing each position’s degrees of freedom from $d$ to $d-1$. However, the model remains invariant under global orthogonal transformations $\mathbf{Z} \mapsto \mathbf{Z}' = \mathbf{Q} \mathbf{Z}$, where $\mathbf{Q} \in \mathbb{R}^{d\times d}$ is orthogonal, representing rotations or reflections that preserve both geodesic distances and inner products on the sphere. Indeed, for transformed positions $\mathbf{z}_i' = \mathbf{Q} \mathbf{z}_i$ and $\mathbf{z}_j' = \mathbf{Q} \mathbf{z}_j$, we have $\|\mathbf{z}_i'\| = \|\mathbf{Q} \mathbf{z}_i\| = \|\mathbf{z}_i\| = 1$, so the unit norm is preserved. Moreover,  
$$
\mathbf{z}_i'^\top \mathbf{z}_j' = (\mathbf{Q} \mathbf{z}_i)^\top (\mathbf{Q} \mathbf{z}_j) = \mathbf{z}_i^\top \mathbf{Q}^\top \mathbf{Q} \mathbf{z}_j = \mathbf{z}_i^\top \mathbf{z}_j,
$$
so all pairwise inner products, and therefore all geodesic distances, are preserved. Consequently, $\mathcal{L}(\alpha, \beta, \mathbf{Z}) = \mathcal{L}(\alpha, \beta, \mathbf{Q} \mathbf{Z})$, and the configuration is identifiable only up to a global rotation or reflection. This does not affect inference about relative positions but requires post-processing (e.g., Procrustes alignment, see \citealt{borg2005modern} for an in-depth discussion) for visualization and interpretation.

\subsection{Model extensions}

Under the Bayesian paradigm, the spherical latent space formulation admits several natural extensions that enhance modeling flexibility, allow richer structural assumptions, and facilitate scalability in more complex settings. Below, we outline representative extensions together with their mathematical specifications.

\paragraph{Mixed effects model.}  
Let $\mathbf{x}_{i,j} \in \mathbb{R}^p$ denote covariates associated with the actor pair $(i,j)$. The linear predictor can be extended to $\eta_{i,j} = \alpha + \boldsymbol{\gamma}^\top \mathbf{x}_{i,j} + \beta \langle \mathbf{z}_i, \mathbf{z}_j \rangle$, where $\boldsymbol{\gamma} \in \mathbb{R}^p$ are regression coefficients. Priors for $\boldsymbol{\gamma}$ include Gaussian for regularization, Laplace for shrinkage, or horseshoe for sparse, high-dimensional settings. This specification enables joint inference on covariate effects and latent geometry.

\paragraph{Weighted inner product model.}  
Analogous to the eigenmodel, the linear predictor can be generalized to $\eta_{i,j} = \alpha + \beta\,\mathbf{z}_i^\top \mathbf{\Lambda} \mathbf{z}_j$, with $\mathbf{\Lambda} = \textsf{diag}(\lambda_1,\ldots,\lambda_{d-1})$, assigns dimension-specific weights. This allows certain hypersphere dimensions to influence tie probabilities more strongly, with Normal or hierarchical shrinkage priors on the diagonal elements of $\mathbf{\Lambda}$ controlling effective rank and mitigating overparameterization.

\paragraph{Finite mixture prior.}  
To capture latent community structure, we assign a finite mixture prior to the latent positions 
$\mathbf{z}_i \mid \xi_i, \bm{\mu}_{\xi_i}, \kappa_{\xi_i} \overset{\text{ind}}{\sim} \textsf{vMF}(\bm{\mu}_{\xi_i}, \kappa_{\xi_i})$, 
with $\xi_i \overset{\text{iid}}{\sim} \textsf{Cat}(\boldsymbol{\pi})$, and $\boldsymbol{\pi} \sim \textsf{Dir}(\boldsymbol{\alpha})$ controlling the cluster weights. This specification induces clustering in the spherical latent space, producing interpretable block structures while preserving the manifold’s geometric constraints.

\paragraph{Dirichlet process prior.}  
A nonparametric alternative replaces the finite mixture with a Dirichlet process (DP) prior, $\mathbf{z}_i \mid G \overset{\text{iid}}{\sim} G$, with $G \sim \textsf{DP}(\alpha_0, G_0)$, where $\alpha_0$ is a concentration parameter and $G_0$ is a suitable base measure on $\mathbb{S}^{d-1}$, such as a $\textsf{vMF}(\bm{\mu}_0, \kappa_0)$ distribution. The DP formulation enables the number of latent clusters to be inferred directly from the data, preserving the geometric constraints of the spherical latent space.

\paragraph{Multilayer networks.}  
For multiplex or multilayer relational data $\mathbf{Y}_1,\ldots,\mathbf{Y}_L$, the linear predictor in layer $\ell$ can be expresed as $\eta_{i,j,\ell} = \alpha_\ell + \beta_\ell \langle \mathbf{z}_i, \mathbf{z}_j \rangle$, with layer-specific intercepts $\alpha_\ell$ and similarity coefficients $\beta_\ell$ assigned hierarchical Normal priors to enable information sharing across layers, thus capturing both layer-specific tie formation mechanisms and cross-layer dependencies.

\paragraph{Dynamic networks.}  
For temporal networks $\mathbf{Y}_1,\ldots,\mathbf{Y}_T$, latent positions $\mathbf{z}_{i,t}$ can be assigned a first-order Markov process on the sphere via $\mathbf{z}_{i,t} \mid \mathbf{z}_{i,t-1} \sim \textsf{vMF}(\mathbf{z}_{i,t-1},\kappa_z)$, while layer-specific intercepts and similarity coefficients evolve as $\alpha_t \mid \alpha_{t-1} \sim \textsf{N}(\alpha_{t-1},\sigma_{\alpha}^2)$ and $\beta_t \mid \beta_{t-1} \sim \textsf{N}(\beta_{t-1},\sigma_{\beta}^2)$, introducing Markovian dependence that allows the model to capture smooth temporal variation in both latent geometry and tie formation tendencies.

\paragraph{Alternative computational strategies.}  
While MCMC is the standard inference approach for latent space models, alternative methods can improve scalability. Variational inference (e.g., \citealt{blei2017variational}) and stochastic variational inference (e.g., \citealt{hoffman2013stochastic}) provide deterministic approximations to the posterior, with stochastic gradient optimization enabling application to large-scale networks.

\section{Inference and computation}\label{section4}

The Euclidean and spherical formulations of latent space models introduced above involve the intercept parameter $\alpha$, the scaling parameter $\beta$, and the matrix of latent positions $\mathbf{Z}$, which together determine the structure of the observed network. Once the model formulation is complete, the next step is to estimate these parameters given the observed network data $\mathbf{Y}$. This section provides a detailed treatment of both classical and Bayesian inference approaches, covering optimization techniques for point estimation as well as sampling algorithms for full posterior inference.

\subsection{Classical inference}

Here, we conduct classical inference via maximum likelihood estimation (e.g., \citealt{casella2024statistical}). In latent space models, the likelihood function is generally not globally concave. This non-concavity complicates classical inference, as multiple local maxima necessitate iterative optimization methods and carefully selected initialization strategies to mitigate the risk of suboptimal convergence. To obtain the maximizer of the log-likelihood, we use gradient descent with a backtracking line search based on the Armijo condition (see \citealt{boyd2004convex} and \citealt{nocedal2006numerical}), ensuring sufficient decrease of the objective at each iteration even in non-concave settings.

Let $\ell(\theta, \mathbf{Z})$ denote the log-likelihood, viewed as a function of the latent positions $\mathbf{Z}$ and the other model parameters collected in $\theta$. Starting from initial values $(\theta^{(0)}, \mathbf{Z}^{(0)})$, the parameters are updated jointly as
\[
(\theta^{(b+1)}, \mathbf{Z}^{(b+1)}) = (\theta^{(b)}, \mathbf{Z}^{(b)}) + r^{(b)} \mathbf{D}^{(b)},
\]
where $\mathbf{D}^{(b)} = \nabla \ell(\theta^{(b)}, \mathbf{Z}^{(b)})$ is the gradient of the log-likelihood with respect to all model parameters, and $r^{(b)}$ is a step size chosen to satisfy the Armijo condition
\[
\ell(\theta^{(b+1)}, \mathbf{Z}^{(b+1)}) \geq \ell(\theta^{(b)}, \mathbf{Z}^{(b)}) + c\, r^{(b)} \nabla \ell(\theta^{(b)}, \mathbf{Z}^{(b)})^\top \mathbf{D}^{(b)},
\]
for some constant $c \in (0,1)$. This ensures a sufficient ascent in the log-likelihood at each step while preserving convergence guarantees.

Because the log-likelihood landscape is highly sensitive to initialization, we examine multiple choices for the starting point $(\theta^{(0)}, \mathbf{Z}^{(0)})$, including random configurations. In the spherical model, optimization is additionally constrained by the manifold geometry, requiring each latent position to remain on the unit sphere. This is enforced by projecting the updated positions back onto the sphere after each gradient step. The procedure is summarized in Algorithm \ref{alg:gradient-armijo}. Given these challenges and the model’s structure, classical inference may not always produce reliable or interpretable estimates. This motivates the use of Bayesian inference, which not only accounts for parameter uncertainty but also incorporates regularization naturally through the prior distribution on the model parameters.

\begin{algorithm}[!htb]
\caption{Gradient ascent with Armijo backtracking line search for maximum likelihood estimation in latent space models}
\label{alg:gradient-armijo}
\begin{algorithmic}[1]
\Require Initial parameters $(\theta^{(0)}, \mathbf{Z}^{(0)})$, step size parameters $\eta > 0$, $\rho \in (0,1)$, Armijo constant $c \in (0,1)$
\State Set $k \gets 0$
\While{not converged}
    \State Compute gradient: $\mathbf{D}^{(b)} \gets \nabla \ell(\theta^{(b)}, \mathbf{Z}^{(b)})$
    \State Initialize step size: $r^{(b)} \gets \eta$
    \While{Armijo condition not satisfied}
        \State Check if $\ell((\theta^{(b)}, \mathbf{Z}^{(b)}) + r^{(b)} \mathbf{D}^{(b)}) \geq \ell(\theta^{(b)}, \mathbf{Z}^{(b)}) + c \, r^{(b)} \, \nabla \ell(\theta^{(b)}, \mathbf{Z}^{(b)})^\top \mathbf{D}^{(b)}$
        \If{condition not satisfied}
            \State $r^{(b)} \gets \rho \, r^{(b)}$
        \EndIf
    \EndWhile
    \State Update parameters: $(\theta^{(b+1)}, \mathbf{Z}^{(b+1)}) \gets (\theta^{(b)},\mathbf{Z}^{(b)}) + r^{(b)} \mathbf{D}^{(b)}$
    \If{spherical model}
        \State Project each row of $\mathbf{Z}^{(b+1)}$ onto the unit sphere
    \EndIf
    \State $k \gets k + 1$
\EndWhile
\State \textbf{Output:} Estimates $(\theta^{\text{ML}}, \mathbf{Z}^{\text{ML}})$
\end{algorithmic}
\end{algorithm}

\subsection{Bayesian inference}

Bayesian inference offers a flexible alternative to traditional statistical learning methods (e.g., \citealt{gelman2013bayesian}), providing both parameter estimation and uncertainty quantification through the full posterior distribution $p(\theta, \mathbf{Z} \mid \mathbf{Y})$. To perform full Bayesian inference, it is first necessary to specify a sensible prior distribution $p(\theta, \mathbf{Z})$. To do so, we adopt independent priors for the latent positions $\mathbf{z}_1, \ldots, \mathbf{z}_n$, chosen to encode plausible assumptions about their spatial configuration before observing the data, and separately assign a prior distribution to the remaining model parameters collected in $\theta$. These priors regularize while respecting the model’s invariances, preventing degeneracy and stabilizing computation.

In the Euclidean case, we assign an isotropic Gaussian prior to the latent positions, $\mathbf{z}_{i} \overset{\text{iid}}{\sim} \mathsf{N} \big(\mathbf{0}, \sigma_{z}^{2}\,\mathbf{I}_{d}\big)$, for $i = 1, \ldots, n$, and independently set 
$\alpha \sim \mathsf{N} \big(\mu_\alpha, \sigma_{\alpha}^{2}\big)$. This prior is rotationally and translationally symmetric, aligning with the likelihood’s invariance under orthogonal transformations. It avoids directional bias, preserves the geometry of the latent space, and softly shrinks positions toward the origin, thereby controlling the unbounded nature of Euclidean embeddings and preventing norm divergence.

In the spherical case, we place a uniform prior on the hypersphere, $\mathbf{z}_{i} \overset{\text{iid}}{\sim} \mathsf{U} \big(\mathbb{S}^{d-1}\big)$, for $i = 1, \ldots, n$, and a correlated Gaussian prior on $(\alpha, \beta)$ with
$\alpha \sim \mathsf{N} \big(\mu_\alpha, \sigma_{\alpha}^2\big)$, $\beta \sim \mathsf{N} \big(\mu_\beta, \sigma_{\beta}^2\big)$, and $\textsf{Cov}(\alpha, \beta) = \rho\,\sigma_{\alpha}\sigma_{\beta}$. The uniform prior corresponds to the uniform measure on $\mathbb{S}^{d-1}$, ensuring invariance under all orthogonal transformations in $\mathbb{R}^d$. This symmetry makes all orientations of the latent configuration equally likely, fully respecting the rotational and reflectional invariance of the spherical latent space.

\begin{figure}[!htb]
    \centering
    \subfigure[Euclidean model.] {\includegraphics[scale=0.4]{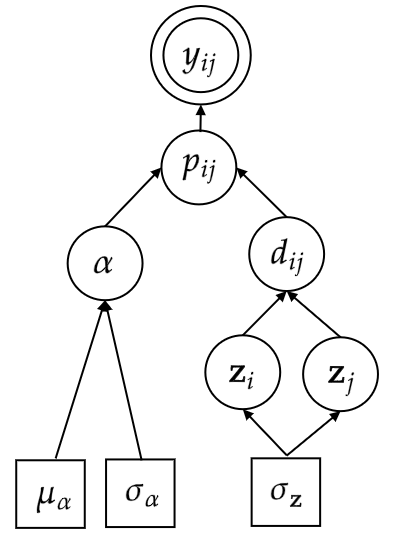}}
    \subfigure[Spherical model.] {\includegraphics[scale=0.4]{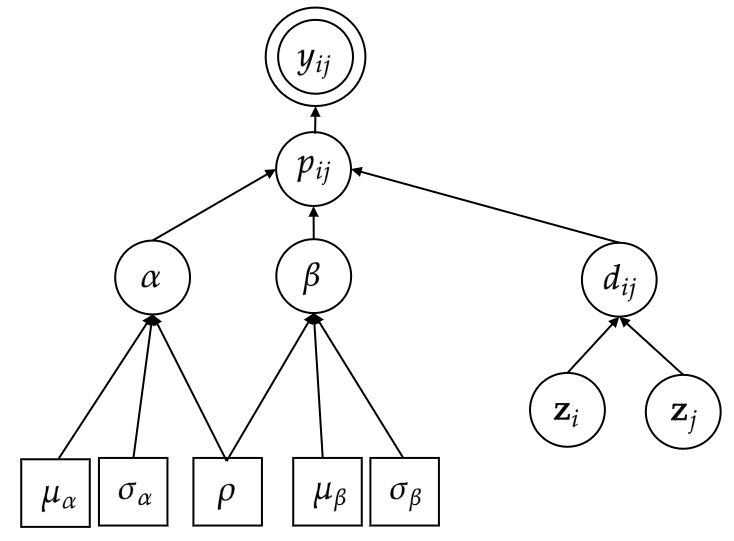}}
    \caption{Directed acyclic graphs (DAGs) for the latent space models, where circles denote random variables or random vectors, squares indicate fixed parameters, and edges represent dependencies.}
    \label{fig:DAG}
\end{figure}

To clarify the probabilistic structure of the models, we present in Figure~\ref{fig:DAG} the directed acyclic graphs (DAGs) corresponding to both the Euclidean and spherical formulations. These graphical representations explicitly illustrate the conditional dependencies among observed ties, latent positions, and global parameters, making clear the conditional independence assumptions underlying the model specification.

Under the prior specifications given above, let the interaction probabilities be defined as $p_{i,j} = \textsf{logit}^{-1}(\eta_{i,j})$, where $\eta_{i,j} = \eta_{i,j}(\theta, \mathbf{z}_i, \mathbf{z}_j) $ is a linear predictor specified as in Section 3, depending on the case. The posterior distribution for each model is proportional to:
\begin{itemize}
    \item Euclidean latent distance model:
    \begin{align*}
        p(\mathbf{Z}, \alpha \mid \boldsymbol{\mathrm{Y}}) &\propto p(\boldsymbol{\mathrm{Y}} \mid \alpha, \mathbf{Z}) \cdot \prod_{i} p(\mathbf{z}_i) \cdot p(\alpha)  \\
        &\propto \prod_{i < j} p_{i,j}^{y_{i,j}} (1 - p_{i,j})^{1-y_{i,j}}   \cdot  \exp\left(-\frac{1}{2\sigma_{z}^2}\textstyle\sum_{i}\|\mathbf{z}_{i}\|^2\right) \cdot
        \exp\left(-\frac{1}{2\sigma_{\alpha}^2}(\alpha-\mu_\alpha)^2\right),
    \end{align*}
    where $\mu_\alpha,\sigma_\alpha^2$, and $\sigma^2_z$ are known hyperparameters.
    \item Spherical latent distance model:
    \begin{align*}
        p(\alpha , \beta, \mathbf{Z} \mid \boldsymbol{\mathrm{Y}}) &\propto p(\boldsymbol{\mathrm{Y}} \mid \alpha,\beta,\mathbf{Z}) \cdot \prod_{i} p(\mathbf{z}_i) \cdot p(\alpha, \beta)  \\
        &= \prod_{i < j} p_{i,j}^{y_{i,j}} (1 - p_{i,j})^{1-y_{i,j}} \cdot \exp\left(-\frac{1}{2(1-\rho^2)} Q(\alpha, \beta) \right),
    \end{align*}
    where $\mu_\alpha,\sigma_\alpha^2,\mu_\beta,\sigma_\beta^2,\rho$, and $\sigma^2_z$ are known hyperparameters, and
    \begin{equation*}
    Q(\alpha, \beta) = \frac{(\alpha - \mu_\alpha)^2}{\sigma_\alpha^2} - \frac{2\rho \, \alpha (\beta - \mu_\beta)}{\sigma_\alpha \sigma_\beta} + \frac{(\beta - \mu_\beta)^2}{\sigma_\beta^2}.    
    \end{equation*}
\end{itemize}

Posterior summaries such as the posterior mean $(\theta^{\text{PM}}, \mathbf{Z}^{\text{PM}}) = \textsf{E}(\theta, \mathbf{Z} \mid \mathbf{Y})$ and the maximum a posteriori $(\theta^{\text{MAP}}, \mathbf{Z}^{\text{MAP}}) = \arg\max p(\theta, \mathbf{Z} \mid \mathbf{Y})$ provide interpretable point estimates of the model parameters, capturing the central tendency of the latent structure, while credible regions derived from the posterior distribution quantify the uncertainty associated with these estimates.

\subsubsection{Sampling algorithms}

The posterior distributions of the latent space models described here are high-dimensional and analytically intractable. To address this challenge, we approximate the posterior using Markov Chain Monte Carlo (MCMC; e.g., \citealt{gamerman2006markov}) methods, specifically the Metropolis–Hastings algorithm (e.g., \citealt{albert2009bayesian}), the Hamiltonian Monte Carlo algorithm (e.g., \citealt{gelman2013bayesian}), and the Spherical Hamiltonian Monte Carlo algorithm (e.g., \citealt{yu2020spherical}), which are developed throughout this section. These approaches provide the flexibility to design geometry-aware proposal distributions tailored to both Euclidean and spherical latent spaces.

\textbf{\textit{Metropolis--Hastings algorithm}}

Let $\mathbf{Z}$ denote the matrix of latent positions, and let $\theta$ represent the global model parameters ($\alpha$ in the Euclidean case, and both $\alpha$ and $\beta$ in the spherical formulation). The unnormalized log-posterior density is  
\[
\psi(\theta, \mathbf{Z}) = \log p(\mathbf{Y} \mid \theta, \mathbf{Z}) + \log p(\mathbf{Z}) + \log p(\theta),
\]  
which is the sum of the log-likelihood and the log-prior densities for the latent positions and the model parameters. The Markov chain is initialized by sampling $\theta$ and $\mathbf{Z}$ from their prior distributions, although maximum likelihood estimates may also serve as effective starting values. We employ a Metropolis algorithm with symmetric proposal distributions, adjusting their variances during the burn-in phase to achieve target acceptance rates, typically between 30\% and 50\%. The joint posterior is explored by alternately updating $\theta$ and $\mathbf{Z}$ via Metropolis steps, with minor procedural differences between the Euclidean and spherical formulations.

Latent positions are updated sequentially by proposing a candidate $\mathbf{z}_i^\ast$, for each $i = 1, \ldots, n$, one at  a time. In the Euclidean case, proposals are drawn from a multivariate Normal distribution centered at the current position, $\mathbf{z}_i^\ast \sim \textsf{N}(\mathbf{z}_i^{(b)}, \tau_{z}\, \mathbf{I}_d)$, where $\tau_{z}$ is a tuning parameter. In the spherical case, following \cite{wood_simulation_1994} and \cite{mardia2000directional}, proposals are drawn from a von Mises--Fisher distribution with mean direction equal to the current position, $\mathbf{z}_i^\ast \sim \textsf{vMF}(\mathbf{z}_i^{(b)}, \tau_{z})$. The proposed value is accepted with probability $\min\big\{ 1, \exp [ \psi(\theta^{(b)}, \mathbf{z}_i^\ast) - \psi(\theta^{(b)}, \mathbf{z}_i^{(b)}) ] \big\}$, where the log-posterior is evaluated holding all latent positions fixed except for node $i$. If accepted, the proposed value becomes the new state $\mathbf{z}_i^{(b+1)} = \mathbf{z}_i^\ast$; otherwise, the position remains unchanged.

The global parameters are updated sequentially. For each parameter, a candidate $\theta^\ast$ is proposed from a Normal distribution centered at the current value, $\theta^\ast \sim \textsf{N}(\theta^{(b)}, \tau_\theta)$, where $\tau_{\theta}$ is once again a tuning parameter. The acceptance probability is computed as $\min\big\{1, \exp[ \psi(\theta^\ast, \mathbf{Z}^{(b)}) - \psi(\theta^{(b)}, \mathbf{Z}^{(b)}) ] \big\}$, analogous to the latent position updates, and the proposed value is accepted or rejected accordingly.

Finally, to account for isometric invariance in the latent space, each sampled configuration $\mathbf{Z}^{(b+1)}$ is aligned to the maximum likelihood estimate $\mathbf{Z}^{\text{ML}}$ using a Procrustes transformation (e.g., \citealt{borg2005modern}). In the Euclidean case, the transformation $\mathbf{T}$ minimizes the trace of the squared distances:
\begin{equation*}\label{classicalProcrutes}
\tilde{\mathbf{Z}}^{(b+1)} = \underset{\mathbf{T}}{\arg\min} \ \text{tr} \left[ \left(\mathbf{Z}^{\text{ML}} - \mathbf{T}\,\mathbf{Z}^{(b+1)}\right)^\top \left(\mathbf{Z}^{\text{ML}} - \mathbf{T}\,\mathbf{Z}^{(b+1)}\right) \right].
\end{equation*}
In the spherical case, the alignment is performed via an orthogonal transformation $\mathbf{T}$ that minimizes the Frobenius norm (e.g., \citealt{golub2013matrix}):
\begin{equation*}\label{ortProcrustes}
\tilde{\mathbf{Z}}^{(b+1)} = \underset{\mathbf{T}}{\arg\min} \ \left\| \mathbf{Z}^{\text{ML}} - \mathbf{T}\,\mathbf{Z}^{(b+1)} \right\|_F,
\end{equation*}
with $\mathbf{T}^\top \mathbf{T} = \mathbf{I}_n$.

After discarding an initial burn-in period and applying thinning when necessary, the retained posterior samples can be used to compute summary statistics of interest, including point estimates and marginal credible intervals for all model parameters, including latent positions. Convergence and mixing are evaluated using standard MCMC diagnostics across multiple chains, such as the effective sample size (ESS) and the potential scale reduction factor $\hat{R}$. The complete estimation procedure is outlined in Algorithm~\ref{alg:MHprocedure}.

\begin{algorithm}[!htb]
\caption{Metropolis--Hastings algorithm for latent space models}
\label{alg:MHprocedure}
\begin{algorithmic}[1]
\Require Observed data $\mathbf{Y}$, priors for $\theta$ and $\mathbf{Z}$, number of iterations $B$
\State Compute $(\theta^{\text{ML}}, \mathbf{Z}^{\text{ML}})$ via optimization using Algorithm~\ref{alg:gradient-armijo}
\State Initialize $(\mathbf{Z}^{(0)}, \theta^{(0)})$ 
\ForAll{chains}
  \For{$b = 1$ \textbf{to} $B$}
    \For{each node $i = 1, \ldots, n$}
      \State Propose $\mathbf{z}_i^\ast$:
      \Statex \hspace{2.5cm} Euclidean: $\mathbf{z}_i^\ast \sim \textsf{N}(\mathbf{z}_i^{(b)}, \tau_{z} \, \mathbf{I}_d)$
      \Statex \hspace{2.5cm} Spherical: $\mathbf{z}_i^\ast \sim \textsf{vMF}(\mathbf{z}_i^{(b)}, \tau_{z})$
      \State Accept $\mathbf{z}_i^\ast$ with probability $\min\big\{ 1, \exp [ \psi(\theta^{(b)}, \mathbf{z}_i^\ast) - \psi(\theta^{(b)}, \mathbf{z}_i^{(b)}) ] \big\}$
      \State Update $\mathbf{z}_i^{(b+1)} \gets \mathbf{z}_i^\ast$ if accepted, else keep $\mathbf{z}_i^{(b)}$
    \EndFor 
    \For{each global parameter $\theta_j$}
      \State Propose $\theta_j^\ast \sim \textsf{N}(\theta_j^{(b)}, \tau_\theta)$
      \State Accept with probability $\min\big\{1, \exp[ \psi(\theta_j^\ast, \mathbf{Z}^{(b)}) - \psi(\theta_j^{(b)}, \mathbf{Z}^{(b)}) ] \big\}$
      \State Update $\theta_j^{(b+1)} \gets \theta_j^\ast$ if accepted, else keep $\theta_j^{(b)}$
    \EndFor
    \State Align $\mathbf{Z}^{(b+1)}$ to $\mathbf{Z}^{\text{ML}}$ using the corresponding transformation
  \EndFor
\EndFor
\State Discard burn-in and apply thinning
\State \textbf{Output:} Chains of posterior samples $(\theta^{(1)},\mathbf{Z}^{(1)}),\ldots,(\theta^{(B)},\mathbf{Z}^{(B)})$
\end{algorithmic}
\end{algorithm}

\textbf{\textit{Hamiltonian Monte Carlo algorithm}}

Alongside the Metropolis--Hastings algorithm, we consider two advanced sampling methods: Hamiltonian Monte Carlo (HMC) and Geodesic Hamiltonian Monte Carlo (GHMC). These approaches exploit gradient information and the underlying geometric structure to explore the posterior distribution more efficiently, particularly in high-dimensional or curved latent spaces. Although they can achieve higher sampling efficiency by accounting for the curvature of the full conditional distributions, they are computationally more demanding because they require gradient evaluations and auxiliary variable generation. Similar to the Metropolis–Hastings algorithm, we design an MCMC scheme in which HMC steps are applied to each latent position $\mathbf{z}_i$, for $i = 1, \ldots, n$, one at a time. For clarity, we present the method in its general form below.

The HMC algorithm expands the parameter space by introducing an auxiliary momentum variable $\mathbf{p}$, which enables the use of Hamiltonian dynamics to generate proposals \citep{neal_mcmc_2011}. For a target unnormalized density $f(\mathbf{x})$, the Hamiltonian is given by
\[
\mathrm{H}(\mathbf{x}, \mathbf{p}) = \mathrm{U}(\mathbf{x}) + \frac{1}{2} \mathbf{p}^\top \mathbf{M}^{-1} \mathbf{p},
\]
where $\mathrm{U}(\mathbf{x}) = -\log f(\mathbf{x})$ represents the potential energy associated with the unnormalized log-density $\log f(\mathbf{x})$, and $\mathbf{M}$ is a symmetric positive-definite mass matrix, often chosen as the identity for simplicity.

The dynamics are determined by Hamilton’s equations and are numerically integrated via the leapfrog method using a fixed step size $\epsilon$ and a predetermined number of steps $L$ \citep{sanz2018numerical}. These algorithmic parameters are tuned during execution to balance computational efficiency and acceptance probability, which typically lies between 60\% and 70\%. A proposal state is obtained by simulating the Hamiltonian dynamics forward in time, after which it is accepted or rejected according to a Metropolis rule that preserves detailed balance. This approach enables the sampler to perform large, gradient-informed moves across the posterior space, offering substantial efficiency gains over regular random-walk proposals \citep{neal_mcmc_2011}. Algorithm~\ref{alg:hmc} outlines the HMC procedure for sampling from a general target distribution.

\begin{algorithm}[!htb]
\caption{Hamiltonian Monte Carlo algorithm}
\label{alg:hmc}
\begin{algorithmic}[1]
\Require Target density $f(\mathbf{x})$, step size $\epsilon$, number of leapfrog steps $L$, mass matrix $\mathbf{M}$
\State Initialize $\mathbf{x}^{(0)}$
\For{$b = 1$ \textbf{to} $B$}
  \State Sample $\mathbf{p}^{(b)} \sim \textsf{N}(\mathbf{0}, \mathbf{M})$
  \State Set $\mathbf{x}^\ast \gets \mathbf{x}^{(b)}$ and $\mathbf{p}^\ast \gets \mathbf{p}^{(b)}$
  \For{each leapfrog step $\ell = 1, \ldots, L$}
    \State $\mathbf{p}^{\ast} \leftarrow \mathbf{p}^{\ast} + \frac{\epsilon}{2} \nabla \log f(\mathbf{x}^{\ast})$
    \State $\mathbf{x}^{\ast} \leftarrow \mathbf{x}^{\ast} + \epsilon\  \mathbf{M}^{-1} \mathbf{p}^{\ast}$
    \State $\mathbf{p}^{\ast} \leftarrow \mathbf{p}^{\ast} + \frac{\epsilon}{2} \nabla \log f(\mathbf{x}^{\ast})$
  \EndFor
  \State Compute $\mathrm{H}_{\text{current}} = -\log f(\mathbf{x}^{(b)}) + \frac{1}{2} \mathbf{p}^{(b)\top} \mathbf{M}^{-1} \mathbf{p}^{(b)}$
  \State Compute $\mathrm{H}_{\text{proposed}} = -\log f(\mathbf{x}^{\ast}) + \frac{1}{2} \mathbf{p}^{\ast\top} \mathbf{M}^{-1} \mathbf{p}^{\ast}$
  \State Accept $\mathbf{x}^{\ast}$ with probability $\min\big\{1, \exp[\mathrm{H}_{\text{current}} - \mathrm{H}_{\text{proposed}}] \big\}$
  \State Update $\mathbf{x}^{(b+1)} \gets \mathbf{x}^{\ast}$ if accepted, else $\mathbf{x}^{(b+1)} \gets \mathbf{x}^{(b)}$
\EndFor
\State Discard burn-in and apply thinning
\State \textbf{Output:} Chains of posterior samples $\mathbf{x}^{(1)},\ldots,\mathbf{x}^{(B)}$
\end{algorithmic}
\end{algorithm}

\textbf{\textit{Spherical Hamiltonian Monte Carlo algorithm}}

Standard Hamiltonian Monte Carlo (HMC) performs effectively in flat Euclidean spaces but can be restrictive when the posterior distribution is supported on a curved manifold. To overcome this limitation, we consider Geodesic Hamiltonian Monte Carlo (GHMC), an extension of HMC designed for embedded Riemannian manifolds, such as spheres or Stiefel manifolds, inspired by the Riemannian Manifold Hamiltonian Monte Carlo (RMHMC) framework of \cite{girolami_rmHMC_2011}. Unlike RMHMC, which requires computing local metric tensors and numerically integrating the geodesic equations, GHMC exploits the manifold’s known geodesic structure to evolve trajectories exactly along curvature-aware paths. In this setting, the target density is expressed with respect to the Hausdorff measure \citep{federer2014geometric}, resulting in the modified Hamiltonian
\[
\mathrm{H}_{\mathcal{H}}(\mathbf{x}, \mathbf{p}) 
= -\log f_{\mathcal{H}}(\mathbf{x}) 
+ \frac{1}{2} \mathbf{p}^\top \mathrm{G}(\mathbf{x})^{-1} \mathbf{p},
\]
where $\mathrm{G}(\mathbf{x})$ is the intrinsic metric tensor of the manifold and $f_{\mathcal{H}}(\mathbf{x})$ is the target density adjusted by the Riemannian volume element $\sqrt{|\mathrm{G}(\mathbf{x})|}$ \citep{byrne_geodesic_2013}.

GHMC alternates between exact geodesic updates of position and velocity (the kinetic flow) and momentum adjustments driven by the gradient of the log-density (the potential flow). On the hypersphere, these geodesic updates admit closed-form solutions involving trigonometric functions of the angular velocity, enabling efficient and numerically stable integration. After a predetermined number of integration steps, a Metropolis acceptance step ensures that the resulting Markov chain targets the correct posterior distribution. To preserve the validity of the dynamics, the momentum is projected onto the tangent space of the manifold after each update. By explicitly exploiting the manifold’s geometric structure, GHMC retains the principal advantages of HMC, such as long-distance proposals and reduced random-walk behavior, while remaining well-defined on curved latent spaces. In contrast to RMHMC, which requires computing and differentiating a position-dependent metric tensor at each iteration \citep{girolami_rmHMC_2011}, GHMC avoids these computational burdens by relying on analytically tractable geodesic flows, making it a scalable alternative for manifold-constrained models \citep{yu2020spherical}. Algorithm~\ref{alg:ghmc} presents the GHMC procedure for sampling from an arbitrary density supported on a spherical manifold.

\begin{algorithm}[!htb]
\caption{Spherical Hamiltonian Monte Carlo algorithm}
\label{alg:ghmc}
\begin{algorithmic}[1]
\Require Target density $f(\mathbf{x})$, step size $\epsilon$, number of leapfrog steps $L$
\State Initialize $\mathbf{x}^{(0)}$
\For{$b = 1$ \textbf{to} $B$}
  \State Sample momentum $\mathbf{p}^{(b)} \sim \textsf{N}(\mathbf{0},\mathbf{I}_d)$
  \State Project onto the tangent space of the manifold $\mathbf{p}^{(b)} \leftarrow (\mathbf{I}_d - \mathbf{x}^{(b)} \mathbf{x}^{(b)\top}) \mathbf{p}^{(b)}$
  \State Set $\mathbf{x}^\ast \gets \mathbf{x}^{(b)}$ and $\mathbf{p}^\ast \gets \mathbf{p}^{(b)}$
  \For{each leapfrog step $\ell = 1, \ldots, L$}
    \State $\mathbf{p}^{\ast} \leftarrow \mathbf{p}^{\ast} + \frac{\epsilon}{2} \nabla \log f(\mathbf{x}^{\ast})$
    \State Project $\mathbf{p}^{\ast} \leftarrow (\mathbf{I}_d - \mathbf{x}^{\ast} \mathbf{x}^{\ast\top}) \mathbf{p}^{\ast}$
    \State Let $\nu \gets \|\mathbf{p}^{\ast}\|$
    \State Geodesic flow update:
    \Statex \hspace{2.5cm} $\mathbf{x}^{\ast} \leftarrow \mathbf{x}^{\ast} \cos(\nu \epsilon) + \frac{1}{\nu}\ \mathbf{p}^{\ast} \sin(\nu \epsilon)$
    \Statex \hspace{2.5cm} $\mathbf{p}^{\ast} \leftarrow \mathbf{p}^{\ast} \cos(\nu \epsilon) - \nu \ \mathbf{x}^{\ast} \sin(\nu \epsilon)$
    \State $\mathbf{p}^{\ast} \leftarrow \mathbf{p}^{\ast} + \frac{\epsilon}{2} \nabla \log f(\mathbf{x}^{\ast})$
    \State Project $\mathbf{p}^{\ast} \leftarrow (\mathbf{I}_d - \mathbf{x}^{\ast} \mathbf{x}^{\ast\top}) \mathbf{p}^{\ast}$
  \EndFor
  \State Compute $\mathrm{H}_{\mathcal{H}\text{ current}} = -\log f(\mathbf{x}^{(b)}) + \frac{1}{2} \mathbf{p}^{(m)\top} \mathbf{p}^{(b)}$
  \State Compute $\mathrm{H}_{\mathcal{H}\text{ proposed}} = -\log f(\mathbf{x}^{\ast}) + \frac{1}{2} \mathbf{p}^{\ast\top} \mathbf{p}^{\ast}$
  \State Accept $\mathbf{x}^{\ast}$ with probability $\min\big\{(1, \exp [ \mathrm{H}_{\mathcal{H}\text{ current}} - \mathrm{H}_{\mathcal{H}\text{ proposed}} ] \big\}$
  \State Update $\mathbf{x}^{(b+1)} \gets \mathbf{x}^{\ast}$ if accepted, else $\mathbf{x}^{(b+1)} \gets \mathbf{x}^{(b)}$
\EndFor
\State Discard burn-in and apply thinning
\State \textbf{Output:} Chains of posterior samples $\mathbf{x}^{(1)},\ldots,\mathbf{x}^{(B)}$
\end{algorithmic}
\end{algorithm}

Hamiltonian-based algorithms must be adapted to latent space network models by tuning hyperparameters, updating parameters sequentially to improve acceptance rates, and respecting the geometry of the latent space. Convergence diagnostics ensure adequate posterior exploration and reliability of inferences, while multimodality and identifiability issues require running multiple chains from diverse initializations and checking the consistency of marginal posteriors. Geometric constraints add further complexity, necessitating careful algorithmic design to balance exploration efficiency, numerical stability, and convergence guarantees.

\subsubsection{Illustration: Rosenbrock function}

In order to illustrate the sampling algorithms given above, we consider sampling from a target distribution of the form $f(\mathbf{x}) \propto \exp(-R(\mathbf{x}))$, where $R(\mathbf{x})$ is the Rosenbrock function \citep{rosenbrock1960} given by:
\begin{equation*}
f(x_1,x_2) = (a-x_{1})^{2} + b(x_{2}-x_{1}^{2})^{2}, \qquad \mathbf{x} \in \mathbb{R}^2 \,\, \text{or} \,\, \mathbb{S}^{1},
\end{equation*}
with parameters $a \in \mathbb{R}$ and $b \in \mathbb{R}$, with $b>0$.

Thus, we apply the Metropolis--Hastings algorithm (Algorithm \ref{alg:MHprocedure}), Hamiltonian Monte Carlo (Algorithm \ref{alg:hmc}), and Geodesic Hamiltonian Monte Carlo (Algorithm \ref{alg:ghmc}). In the experiments, the Rosenbrock parameters are set to $a = 1.0$ and $b = 5.0$. For the Euclidean case $\mathbb{R}^2$, we use initial state $\mathbf{x}^{(0)}=(0.0,0.0)$, $5000$ samples, step size $0.05$, burn-in $50000$, thinning $100$, and $2$ chains. For the circular manifold $\mathbb{S}^1$, we use initial state $\mathbf{x}^{(0)}=(1.0,0.0)$ with the same sampling settings. These configurations ensured adequate exploration of the target distributions while keeping computational cost reasonable (full implementation details are available at \url{https://github.com/cnosa/LatentSpaces_Network_Manifold}).

\begin{figure}[!htb]
    \centering
    \subfigure[MH sampling on~$\mathbb{R}^2$,]{\includegraphics[scale=0.18]{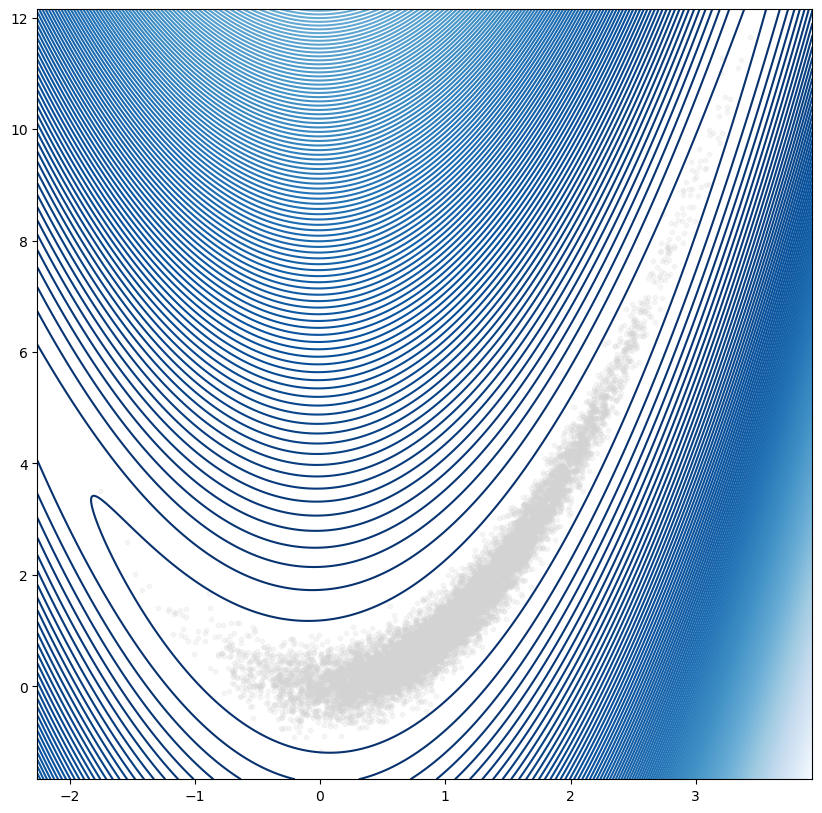}} \hspace{0.1cm}
    \subfigure[HMC sampling on~$\mathbb{R}^2$.]{\includegraphics[scale=0.18]{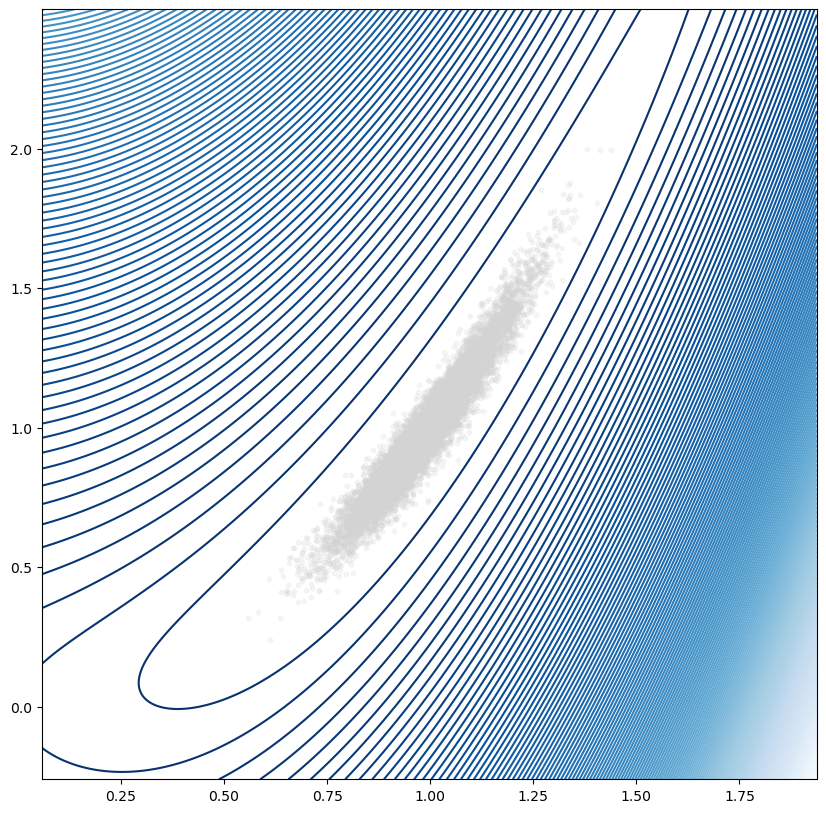}} \hspace{0.1cm}
    \subfigure[MH sampling on~$\mathbb{S}^1$.]{\includegraphics[scale=0.18]{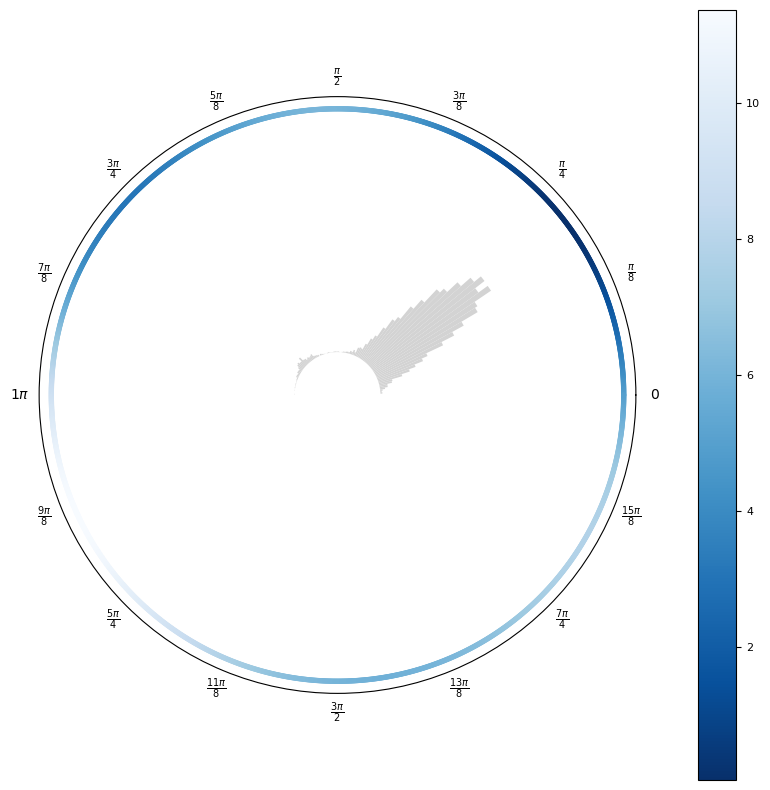}} \hspace{0.1cm}
    \subfigure[GHMC sampling on~$\mathbb{S}^1$.]{\includegraphics[scale=0.18]{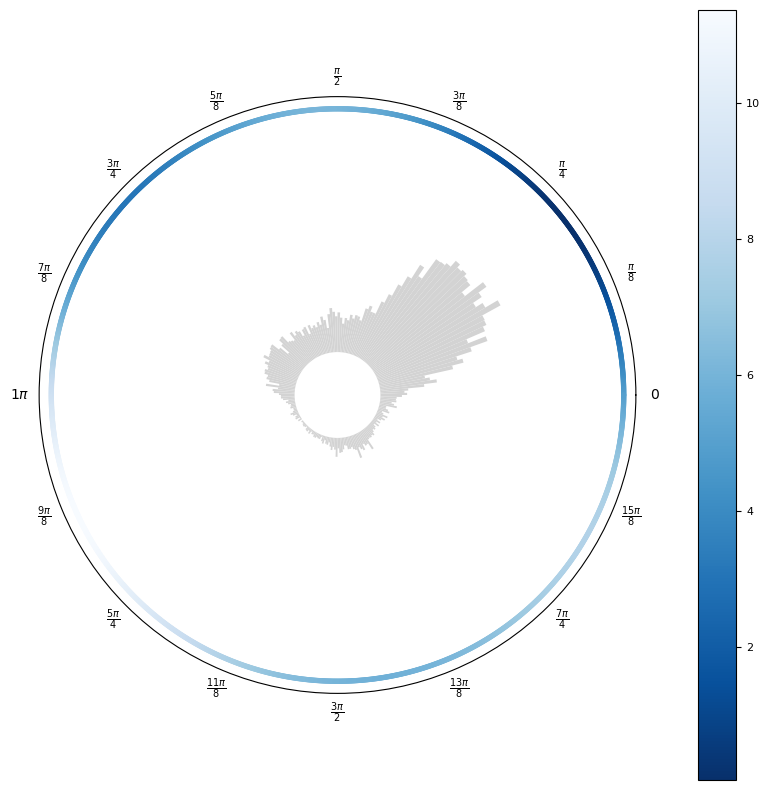}}
    \caption{Comparison of sampling results for the Rosenbrock distribution across different spaces and algorithms.}
    \label{fig:Rosenbrock}   
\end{figure}

The sampling results for the Rosenbrock distribution are presented in Figure~\ref{fig:Rosenbrock} and Table~\ref{tab:Rosenbrock_comparison}. Panels (a) and (b) show the contours of the Rosenbrock function and the sample distributions in $\mathbb{R}^2$ obtained with the MH and HMC algorithms, respectively. HMC more effectively explores the narrow curved valley of the Rosenbrock function, yielding samples with lower variance than MH, which exhibits broader dispersion. Panels (c) and (d) display the Rosenbrock function evaluated on the circle and the corresponding samples under the constraint $\mathbb{S}^1$. Here, MH produces a more limited exploration than Geodesic HMC, reflected in lower sample variance; in both cases, the estimates reveal a bimodal distribution along the circumference.

Table~\ref{tab:Rosenbrock_comparison} reports quantitative performance metrics, including average execution time, maximum $\hat{R}$, minimum relative effective sample size (ESS), average acceptance rate, mean value of the Rosenbrock function, and the tuned step size. HMC consistently achieves higher acceptance rates and larger ESS, indicating better convergence and more efficient exploration of the target distribution. In contrast, MH shows lower acceptance rates and reduced ESS, reflecting slower mixing, but with substantially lower computational cost, as evidenced by shorter execution times.

\begin{table}[!htb]
\centering
\caption{Comparison of Metropolis--Hastings (MH) and Hamiltonian Monte Carlo (HMC) algorithms for sampling from the Rosenbrock distribution on $\mathbb{R}^2$ and $\mathbb{S}^1$. }
\begin{tabular}{ccccc}
\hline
Model                             & \multicolumn{2}{c}{$\mathbb{R}^2$} & \multicolumn{2}{c}{$\mathbb{S}^1$} \\ \hline
Sampling algorithm                & MH              & HMC              & MH                 & GHMC           \\ \hline
Execution time (seconds)          & 40.2300         & 903.9240         & 98.2749            & 27363.6318             \\
Maximum $\hat{R}$                 & 1.0001          & 1.0003           & 1.0001             & 1.0090             \\ 
Minimum relative ESS              & 0.1931          & 0.9999           & 0.9425             & 0.9482             \\ 
Acceptance rate                   & 0.3634          & 0.9340           & 0.1250             & 0.6819             \\ 
Mean value of Rosenbrock function & 0.9989          & 0.0556           & 0.8094             & 3.6834             \\ 
Step size after tunning           & 0.4292          & 0.1567           & 0.0001             & 0.0004             \\ \hline
\end{tabular}
\label{tab:Rosenbrock_comparison}
\end{table}

\section{Illustration: Florentine families}\label{section5}

We illustrate the latent space models using two benchmark social science networks originally analyzed by \cite{hoff_latent_2002}. The main text presents results for the Florentine families data, while the supplementary material reports findings for the monks data.
These datasets differ in scale, domain, and type of interaction, offering a diverse testbed for assessing model behavior and interpretability. For each, we fit the proposed model to the observed relational data and analyze the latent structure, inference results, and model fit via posterior summaries and predictive checks. The examples include a historical marriage network among Florentine families and interpersonal ties among monks in a monastery.

In historical social network analysis, relationships between influential families reveal power dynamics, alliances, and social structures. A notable example is the study by \cite{padgett_robust_1993}, which examines marriage and business ties among 16 prominent Florentine families in the 15th century, drawing on historical accounts that document the socio-political landscape of Renaissance Florence.

\begin{figure}[!htb]
    \centering
    \subfigure[Data with family names.]    {\includegraphics[scale=0.33]{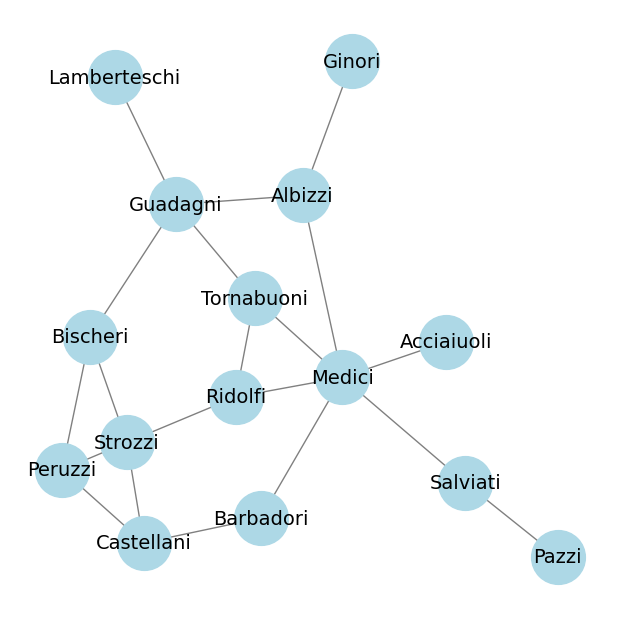}}
    \subfigure[Data with labeled nodes.]   {\includegraphics[scale=0.33]{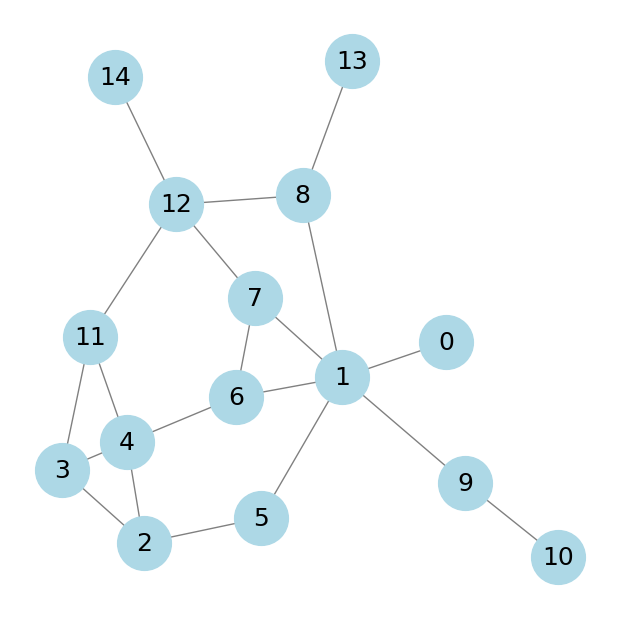}}
    \subfigure[Adjacency matrix.]          {\includegraphics[scale=0.48]{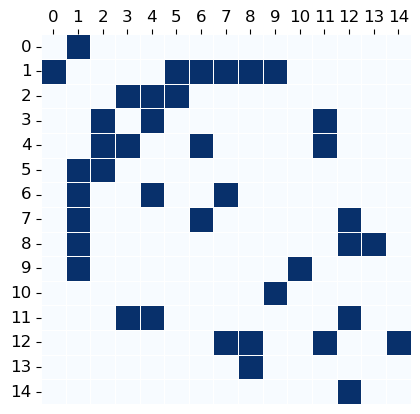}}
    \caption{Visualization of the marriage network among Florentine families.}
    \label{fig:Florentine_network}
\end{figure}

This model focuses on marriage relations between these families, where a tie indicates at least one marriage between them. The network is undirected. One of the 16 families, having no marriage ties, was excluded to avoid infinite distances in maximum likelihood estimation and large finite distances in a Bayesian framework. Modeling these relations with network analysis techniques reveals the structural properties of the Florentine elite and their strategic matrimonial alliances, offering insights into how marriage shaped political and economic power during the Renaissance. Figure~\ref{fig:Florentine_network} shows the network, its relabeling, and the sociomatrix.

We fit the models in $\mathbb{R}^1$, $\mathbb{R}^2$, $\mathbb{R}^3$, $\mathbb{S}^1$, and $\mathbb{S}^2$ using Algorithms~\ref{alg:MHprocedure}, \ref{alg:hmc}, and \ref{alg:ghmc}, with hyperparameters $\sigma_{\mathbf{z}} = 5.0$ and $\sigma_\alpha = 5.0$ for the Euclidean model, and $\rho = -0.5$, $\mu_\alpha = 0.0$, $\sigma_\alpha = 1.0$, $\mu_\beta = 10.0$, and $\sigma_{\beta} = 5.0$ for the spherical model. Hyperparameters were selected via sensitivity analysis to optimize information and predictive criteria (supplementary material). While Hamiltonian dynamics methods (Algorithms~\ref{alg:hmc} and \ref{alg:ghmc}) yield higher effective sample sizes (ESS) per iteration, they require gradient evaluations that increase computation time and may introduce numerical instabilities.

\begin{figure}[!htb]
    \centering
    \includegraphics[scale=0.5]{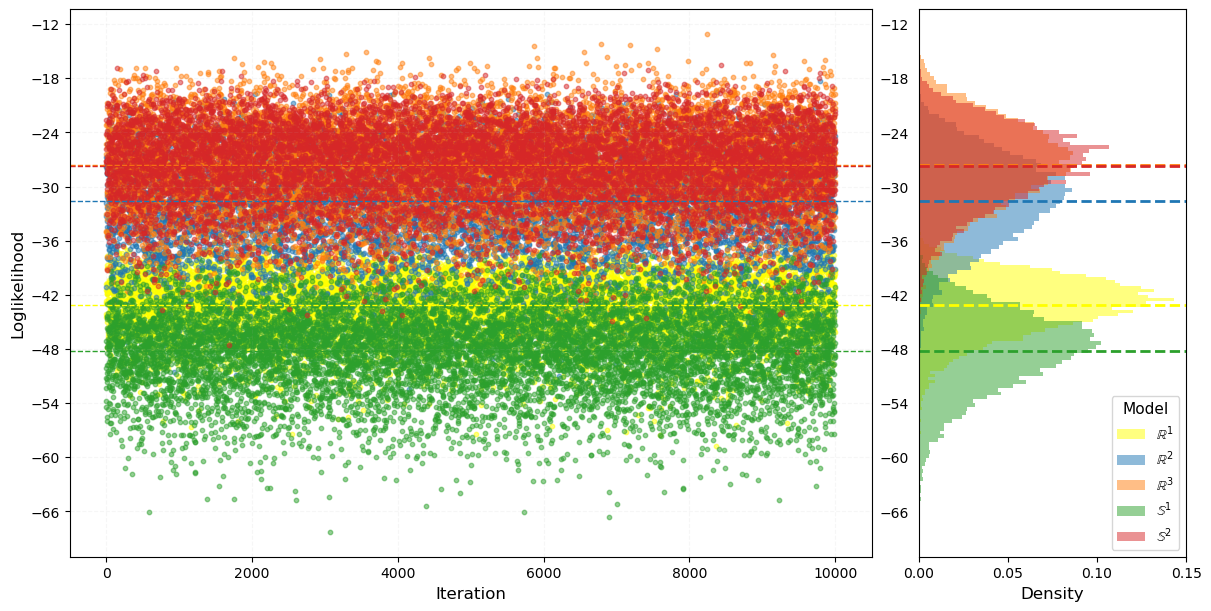}
    \caption{Log-likelihood trajectories for all models obtained during MCMC sampling.}
    \label{fig:Florentine_loglikelihood}
\end{figure}

We use the Metropolis--Hastings (MH) algorithm for the reported results, confirming convergence through standard diagnostics \citep{gelman2013bayesian}. We first assess model fit and performance by inspecting log-likelihood trajectories from MH sampling (Figure~\ref{fig:Florentine_loglikelihood}). All models exhibit good convergence, with the Euclidean model in $\mathbb{R}^3$ achieving the highest likelihood, followed by the spherical model in $\mathbb{S}^2$.

\begin{table}[!htb]
\centering
\begin{tabular}{lccccc}
\hline
& $\mathbb{R}^1$ & $\mathbb{R}^2$ & $\mathbb{R}^3$ & $\mathbb{S}^1$ & $\mathbb{S}^2$ \\
\hline
WAIC           & 207.860 & 170.072 & \textbf{161.099} & 384.106 & 199.367 \\
Mean log-lik.  & -43.105 & -31.639 & \textbf{-27.642} & -48.244 & -27.692 \\
ML log-lik.    & -47.632 & -29.819 & \textbf{-26.517} & -40.270 & -31.852 \\
MAP log-lik.   & -48.392  &-37.978 & -33.1389 & -46.553 & \textbf{-16.570} \\
CM log-lik.    & -51.085 & -36.771 & -26.935 & -45.115 & \textbf{-13.469} \\
\hline
\end{tabular}
\caption{Model comparison metrics for each latent space specification. The mean log-likelihood is averaged over posterior samples, while ML, MAP, and CM denote the log-likelihood evaluated at the maximum likelihood estimate, the maximum a posteriori estimate, and the posterior mean, respectively.}
\label{tab:Florentine_model_comparison}
\end{table}

Table~\ref{tab:Florentine_model_comparison} reports the Watanabe–Akaike information criteria (e.g., \citealt{watanabe2013book}, \citealt{gelman_understanding_2013}) for each model. Spherical models achieve log-likelihood values close to their Euclidean counterparts, indicating comparable ability to capture network structure despite geometric constraints that reduce the effective degrees of freedom: for $n$ nodes and dimension $d$, a model in $\mathbb{R}^d$ has $n d + 1$ degrees of freedom, whereas a model in $\mathbb{S}^{d-1}$ has $n (d - 1) + 2$ due to constraints on latent positions.

\begin{figure}[!b]
    \centering
    \subfigure[Latent space $\mathbb{R}^1$.]{\includegraphics[scale=0.35]{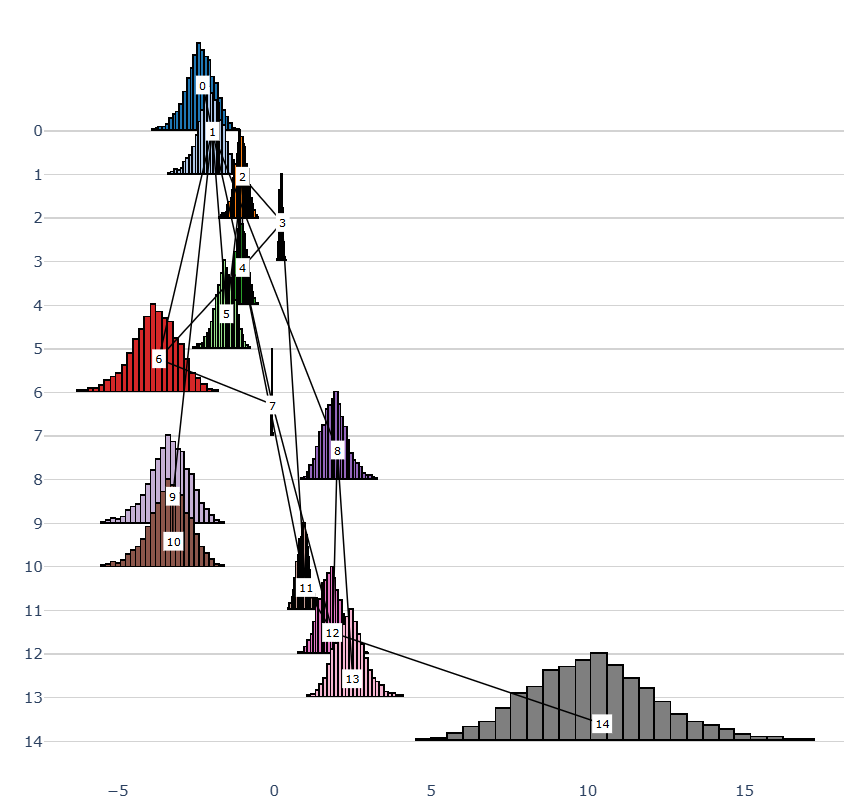}} \hspace{1cm}
    \subfigure[Latent space $\mathbb{S}^1$.]{\includegraphics[scale=0.18]{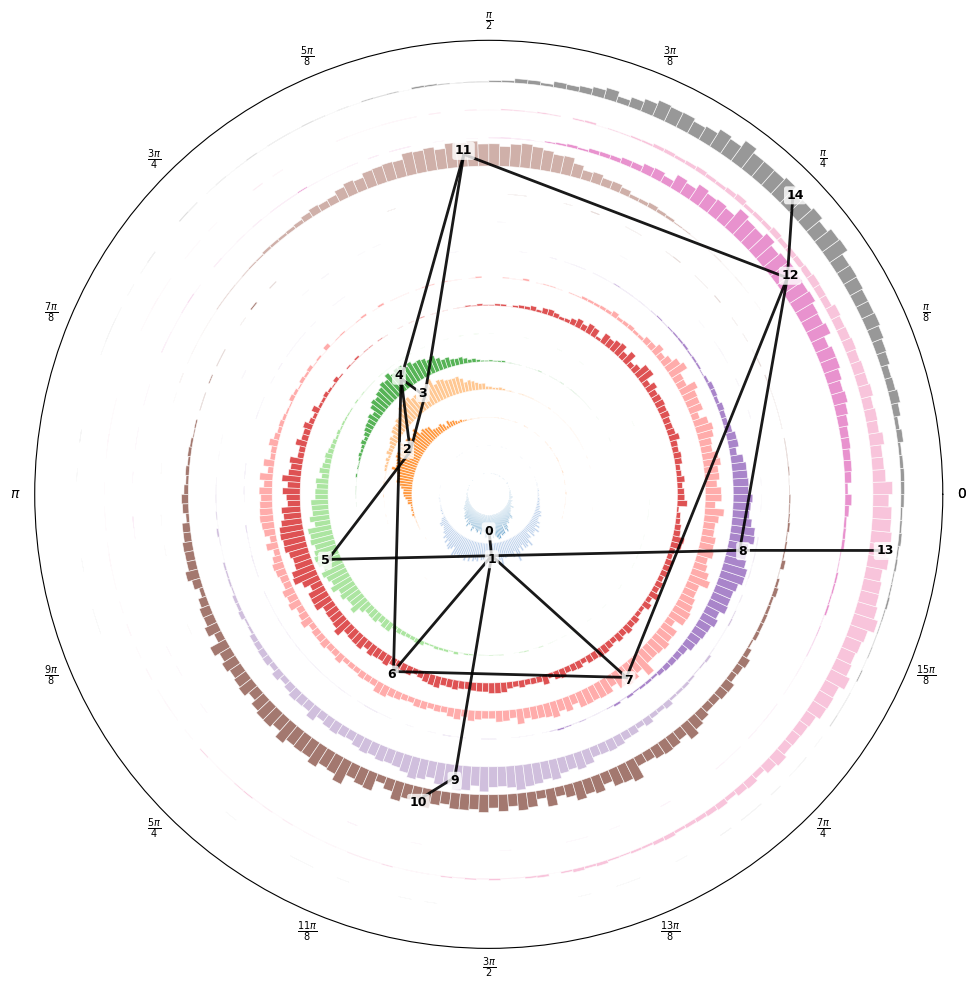}} \\
    \subfigure[Latent space $\mathbb{R}^2$.]{\includegraphics[scale=0.18]{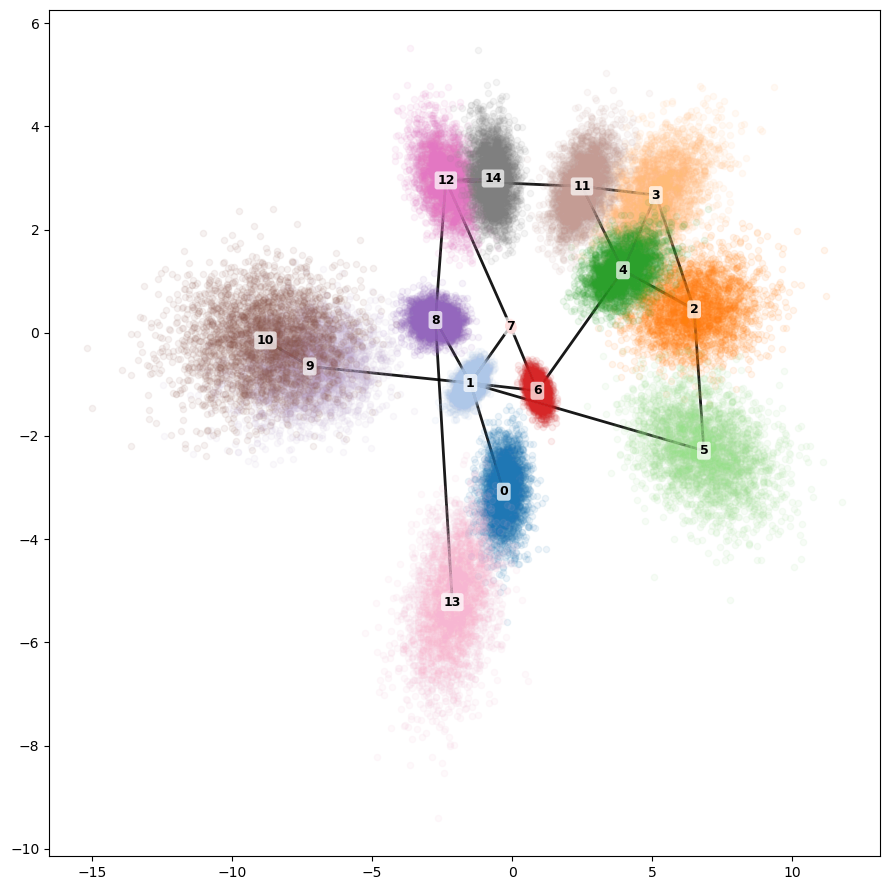}} \hspace{1cm}
    \subfigure[Latent space $\mathbb{S}^2$.]{\includegraphics[scale=0.375]{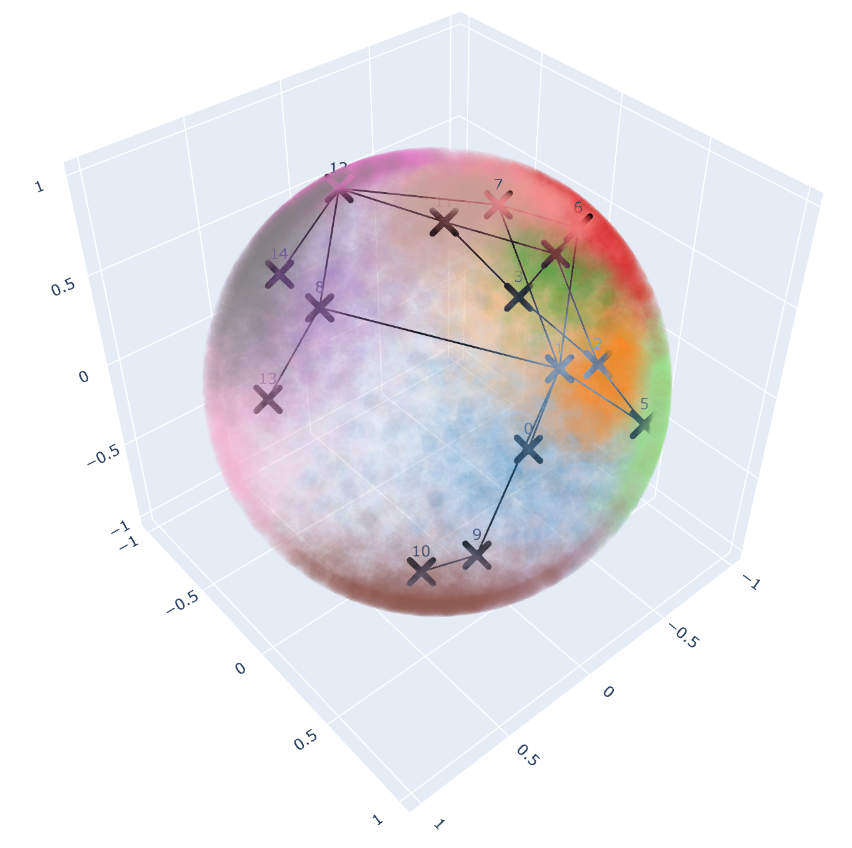}}
    \caption{Latent space representations for $\mathbb{R}^1$ and $\mathbb{S}^1$ models, showing each node’s posterior distribution and mean position, with edges indicating ties from the original network.}
    \label{fig:Florentine_latentspace}   
\end{figure}

Figure~\ref{fig:Florentine_latentspace} shows the estimated node positions for the models in $\mathbb{R}^1$ and $\mathbb{S}^2$. Both Euclidean and spherical embeddings preserve much of the original graph structure, but the circular layout provides a more compact configuration with clearer group separation, potentially reflecting structural features such as hierarchical clustering or cyclical subgroups. Average posterior variance was highest for $\mathbb{R}^3$ (3.157), followed by $\mathbb{R}^2$ (0.987) and $\mathbb{R}^1$ (0.429), with spherical models yielding the tightest configurations $\mathbb{S}^1$ (0.393) and $\mathbb{S}^2$ (0.154).

In the inferred latent spaces, certain families consistently occupy central positions, most notably the Medici, who appear close to many others, indicating high latent connectivity. This agrees with historical accounts \citep{padgett_robust_1993} highlighting their strategic marriage alliances and political integration (node~1). In contrast, families such as the Strozzi (node~4) and Albizzi (node~8) appear more peripheral, reflecting fewer matrimonial ties or deliberate exclusionary practices. In latent space models, proximity directly influences tie probability, so central positions correspond to latent social capital, revealing each family’s embeddedness in the marriage economy of Renaissance Florence. These patterns motivate examining node centrality and assessing the model’s ability to recover historically documented alliances and clustering structures.

To quantify network centrality, we use degree and closeness from classical network theory \citep{kolaczyk_statistical_2020}, along with three latent-space-based measures \citep{sosa_betancourt_multilayer_2022}. The probability connection of node $i$ is the expected average probability of connection to other nodes,
$\textsf{E}[\,\sum_{j \neq i} p_{ij} \mid \mathbf{Y}\,]$,
where higher values indicate denser latent regions. The mean distance of node $i$ is the expected average Euclidean or geodesic distance to other nodes,
$\textsf{E} [\, \sum_{j \neq i} h(\mathbf{z}_{i},\mathbf{z}_{j}) \mid \mathbf{Y} \,]$,
with $h(\mathbf{u},\mathbf{v}) = \|\mathbf{u}-\mathbf{v}\|$ in $\mathbb{R}^d$ and $h(\mathbf{u},\mathbf{v}) = \arccos(\mathbf{u}^\top \mathbf{v})$ in $\mathbb{S}^{d-1}$. The center distance of node $i$ measures the expected distance to the latent space centroid $\bar{\mathbf{z}}$,
$\textsf{E} [\, h(\mathbf{z}_{i},\bar{\mathbf{z}}) \mid \mathbf{Y} \,]$,
where $\bar{\mathbf{z}} = \arg \min_{\mathbf{z}} \sum_{j} h(\mathbf{z}_{j}, \mathbf{z})^2$.
These latent space-based measures offer a probabilistic and geometric view of centrality that complements traditional metrics.

Table~\ref{tab:Florentine_correlation} shows that Euclidean models yield higher correlations with node degree, indicating stronger alignment between spatial proximity and structural centrality. Across geometries, nodes with high degree generally exhibit higher connection probability and lower center distance in Euclidean spaces, reinforcing their spatial coherence. Spherical models display greater variability and weaker alignment, likely due to geometric constraints. Certain nodes (e.g., 1 and 7) remain central across all metrics, underscoring their consistent prominence. These patterns are evident in the latent space visualizations (Figure~\ref{fig:Florentine_latentspace}), where central nodes cluster near core regions.

\begin{table}[!htb]
\centering
\begin{tabular}{lccccc}
\hline
Model                  & $\mathbb{R}^1$ & $\mathbb{R}^2$ & $\mathbb{R}^3$ & $\mathbb{S}^1$ & $\mathbb{S}^2$ \\ \hline
Probability connection & 0.434 & 0.539  &  0.751  &  0.318  &  \textbf{0.754}  \\
Mean distance          & -0.400 & -0.483  & \textbf{ -0.698}  &  -0.258  &  -0.658  \\
Center distance        & -0.432 & -0.472  &  -0.692  &  -0.031  &  \textbf{-0.817}  \\ \hline
\end{tabular}
\caption{Correlation between latent space centrality measures and node degree.}
\label{tab:Florentine_correlation}
\end{table}

\begin{figure}[!htb]
    \centering
    \subfigure[Greedy: $K=3$, $M = 0.185$.] {\includegraphics[scale=0.2]{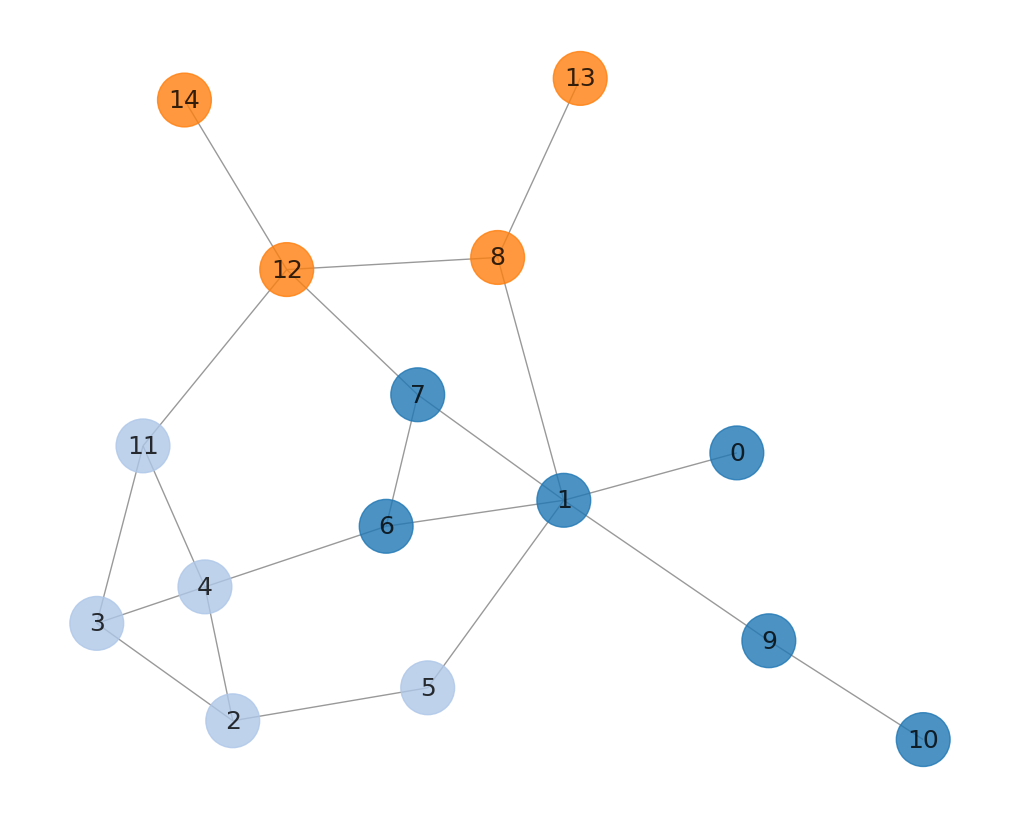}}
    \subfigure[$\mathbb{R}^1$: $K=3$, $M = 0.185$.] {\includegraphics[scale=0.2]{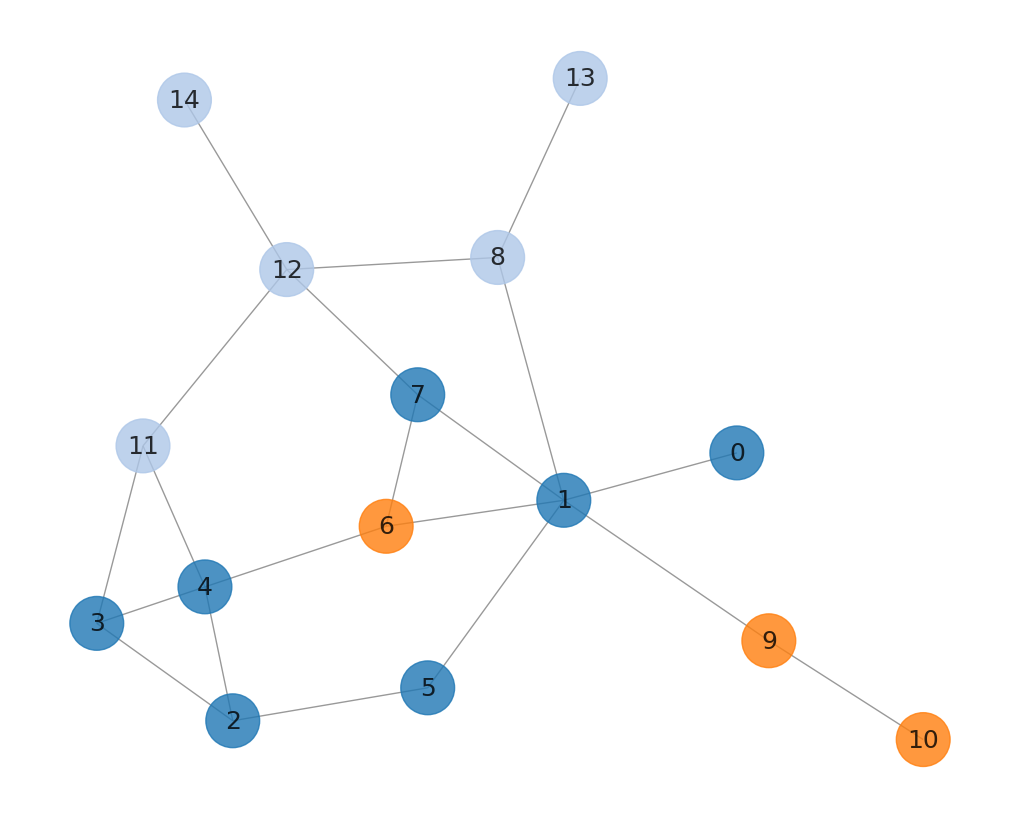}}
    \subfigure[$\mathbb{R}^2$: $K=3$, $M = 0.351$.] {\includegraphics[scale=0.2]{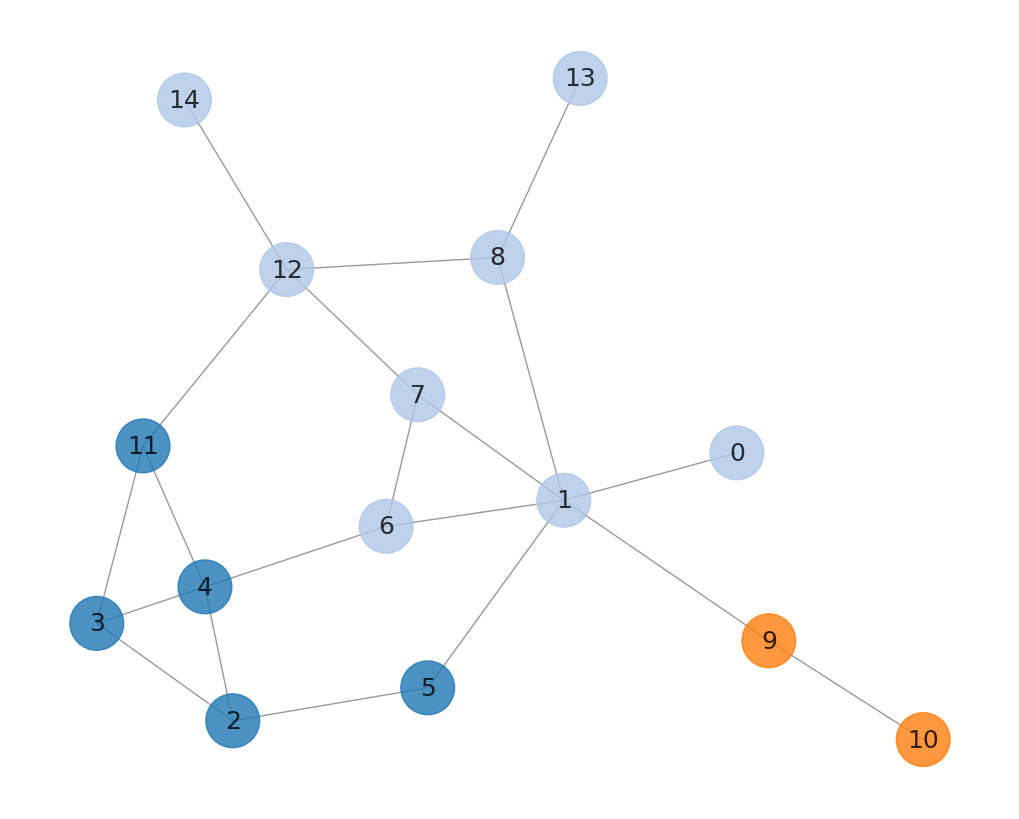}}
    \subfigure[$\mathbb{R}^3$: $K=3$, $M = 0.330$.] {\includegraphics[scale=0.2]{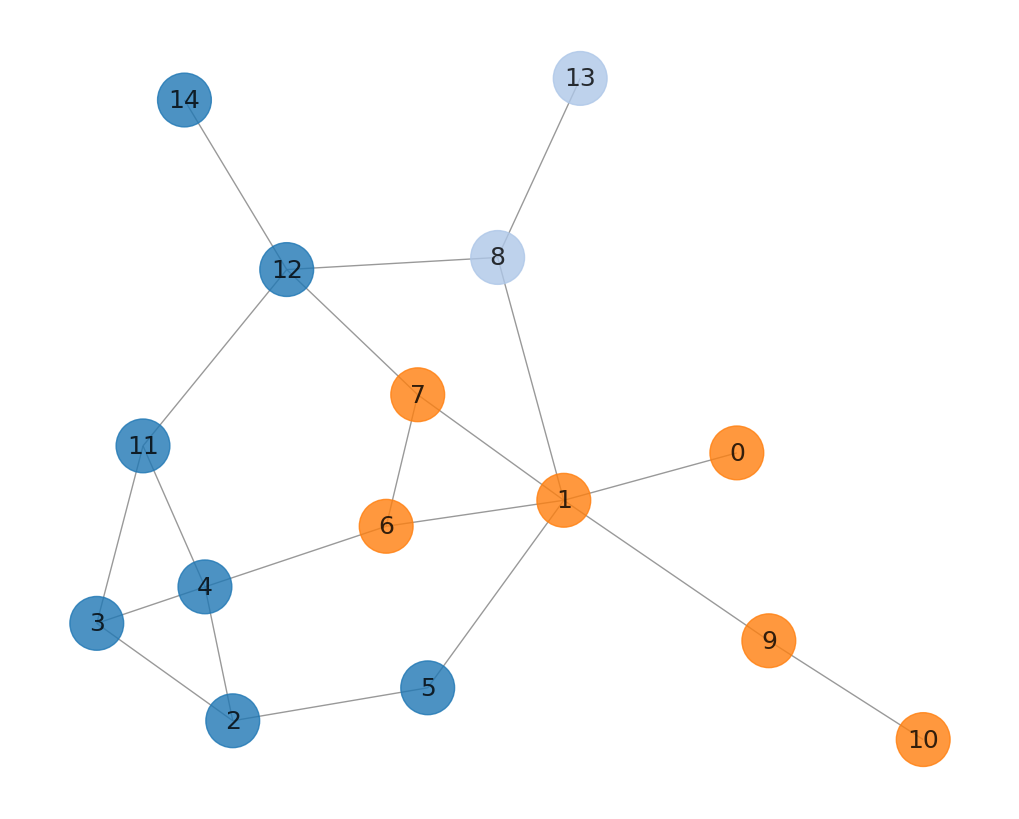}}
    \subfigure[$\mathbb{S}^1$: $K=3$, $M = 0.391$.] {\includegraphics[scale=0.2]{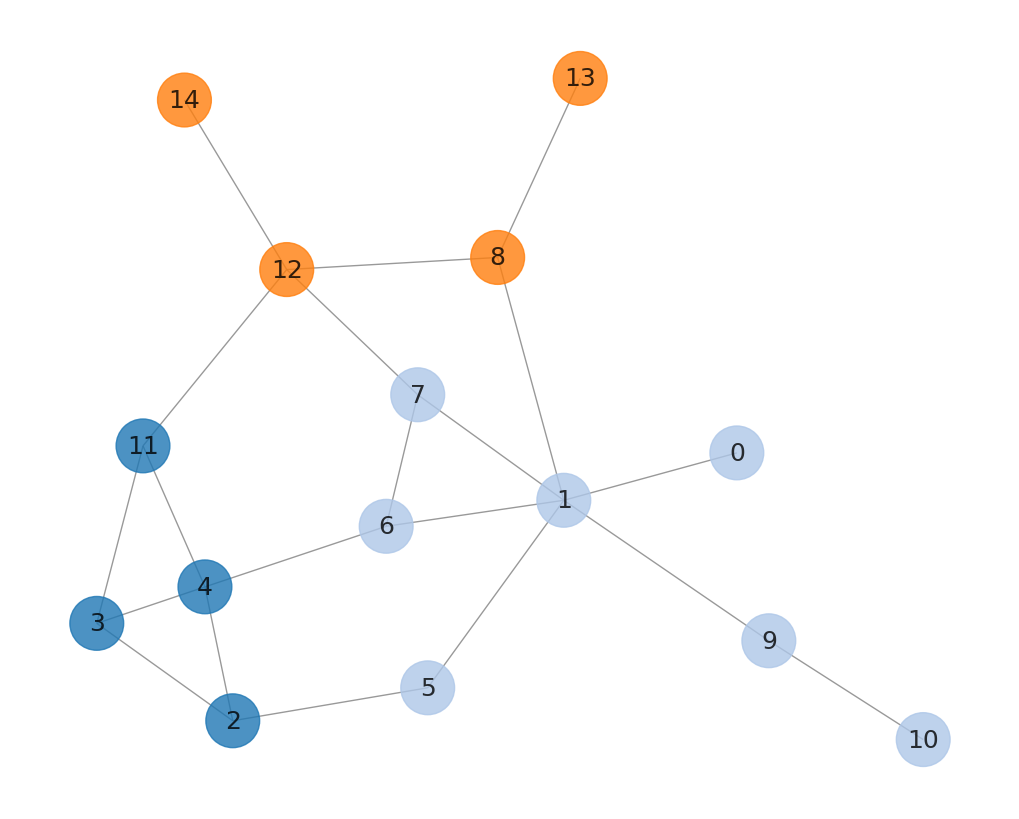}}
    \subfigure[$\mathbb{S}^2$: $K=3$, $M = 0.360$.] {\includegraphics[scale=0.2]{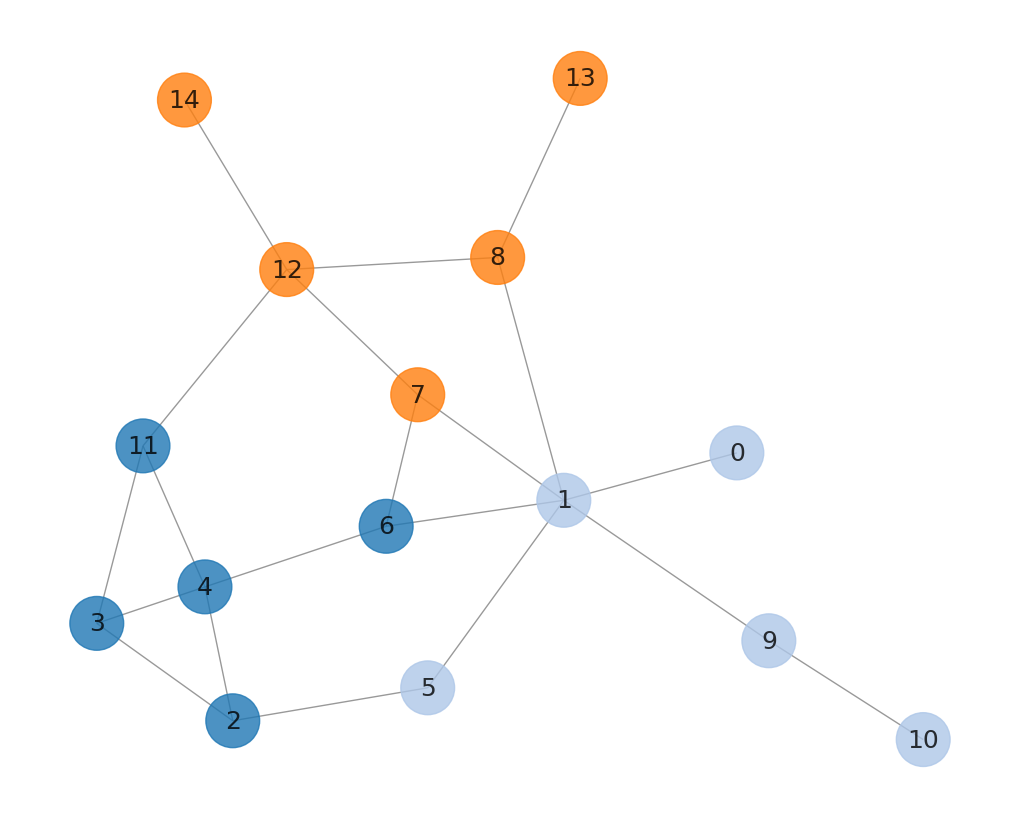}}
    \caption{Clustering across latent space representations, showing number of communities $K$ and modularity $M$.}
    \label{fig:Florentine_clustering}
\end{figure}

Once central nodes are identified, we detect communities by clustering each latent space. Point estimates of latent positions are obtained and used in spectral clustering \citep{ng2002spectral} with geometry-consistent affinity matrices. The number of clusters is chosen via the silhouette score, and the partition with maximum modularity is taken as the representative structure. Figure~\ref{fig:Florentine_clustering} shows the detected communities, their counts, and modularity scores. The $\mathbb{S}^1$ model achieves the highest modularity, indicating the best capture of the network’s community structure.

\begin{figure}[!htb]
    \centering
    \includegraphics[scale=0.26]{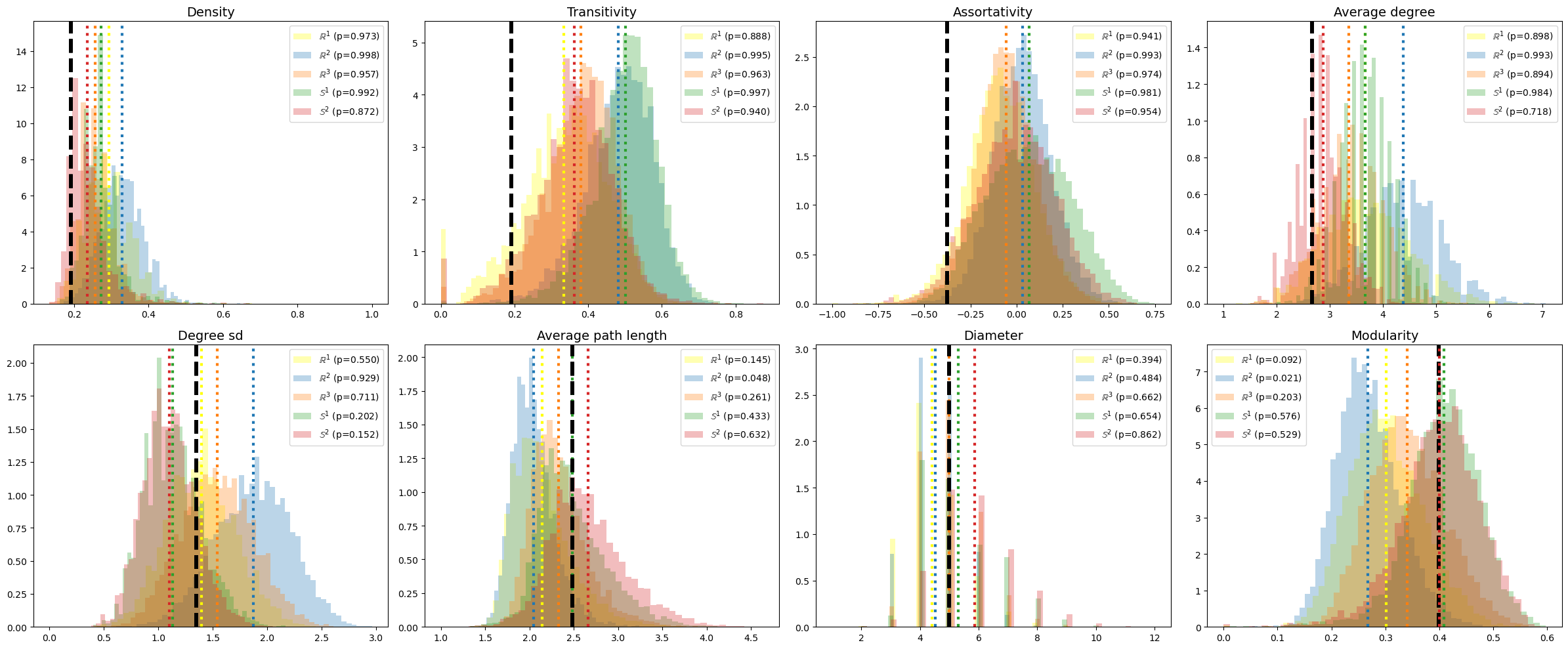}
    \caption{Posterior predictive checks for all models using network-level statistics in the Florentine families example. Each plot shows model-based histograms with means (dashed line) and the observed value (black line), with the legend reporting the posterior predictive $p$-value for each model.}
    \label{fig:Florentine_posteriorpredictivecheck}
\end{figure}

Model adequacy was evaluated through posterior predictive checks using key network-level statistics: degree distribution, geodesic distances, shared edges, and triad census (Figure~\ref{fig:Florentine_posteriorpredictivecheck}). In this network, latent space models primarily capture degree and geodesic distance patterns, as evidenced by the posterior predictive $p$-values and the inclusion of observed statistics within the posterior predictive intervals \citep{gelman_understanding_2013}. For statistics requiring connectedness, computations use the largest connected component of simulated networks drawn from the posterior. Modularity is computed for both observed and simulated networks using the Clauset–Newman–Moore greedy maximization algorithm \citep{clauset2004finding}.

\begin{figure}[!htb]
  \centering
  \begin{minipage}[t]{0.58\textwidth}
    \vspace{-3.5cm}
    \centering
    \begin{tabular}{lccccc}
    \hline
    Model    & $\mathbb{R}^1$  & $\mathbb{R}^2$    & $\mathbb{R}^2$    & $\mathbb{S}^1$    & $\mathbb{S}^2$    \\ \hline
    AUC      & 0.773 & 0.910 & 0.913 & 0.941 & \textbf{0.994} \\ \hline
    Accuracy & 0.780 & 0.819 & 0.933 & 0.923 & \textbf{0.952} \\ \hline
    F1-score & 0.439 & 0.627 & 0.820 & 0.789 & \textbf{0.857} \\ \hline
    \end{tabular}
    \captionof{table}{Predictive performance metrics for each model.}
    \label{tab:Florentine_prediction}
  \end{minipage}
  \hfill
  \begin{minipage}[t]{0.38\textwidth}
    \centering
    \includegraphics[width=\linewidth]{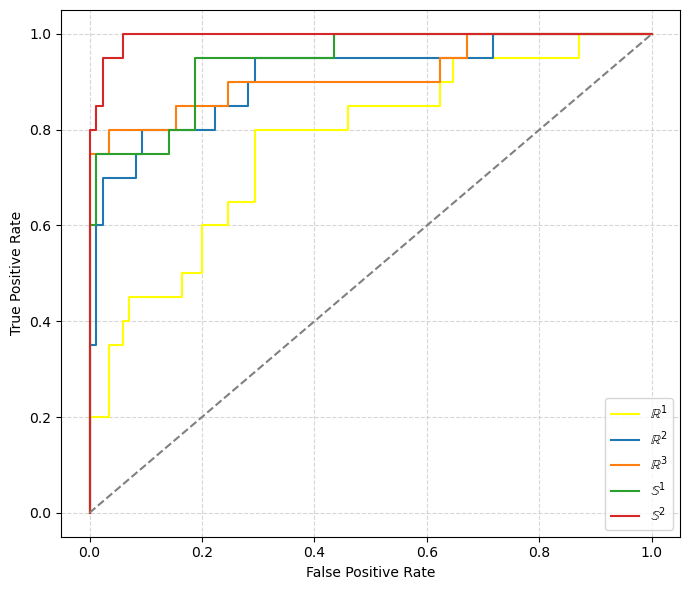}
    \caption{Receiver operating characteristic (ROC) curves for each model.}
    \label{fig:Florentine_ROC}
  \end{minipage}%
\end{figure}

Finally, predictive performance was assessed using out-of-sample AUC, accuracy, and F1-score (Figure~\ref{fig:Florentine_ROC} and Table~\ref{tab:Florentine_prediction}). The $\mathbb{S}^2$ model outperforms all others across metrics, with both spherical models surpassing their Euclidean counterparts, highlighting the role of latent geometry in modeling relational data. Gains in AUC and F1-score indicate improved discrimination and class balance in link prediction. The advantage of $\mathbb{S}^2$ likely stems from its ability to capture transitivity and clustering more effectively than Euclidean spaces, reflecting the dense, alliance-driven structure of Florentine marriage ties.

\section{Discussion}

This paper introduces a spherical latent space model for social network analysis, extending \cite{hoff_latent_2002} by embedding nodes on a hypersphere instead of in Euclidean space. This geometry naturally captures transitivity, directional clustering, and cyclic structures while avoiding degeneracies such as unbounded distances. The approach reflects a broader trend in statistics and machine learning toward manifold-based modeling, that provide intrinsic structures for representing hierarchies, cycles, symmetries, and other topological constraints difficult to model in Euclidean settings. Within this framework, latent space models on manifolds offer principled representations of network geometry, enabling faithful structural characterization and interpretable visualizations.

Spherical latent spaces naturally encode relationships with topological constraints such as cyclicity or periodicity, relevant to settings from social networks to recurring spatial processes. By mapping unobserved relationships to positions on the manifold, they support probabilistic inference and interpretable visualizations of complex structures. However, manifold embeddings constrain the parameter space, may impose unnecessary geometric restrictions, and increase computational cost, especially in Bayesian settings with Hamiltonian or geodesic dynamics. In these contexts, directional statistics are essential, redefining concepts like mean, variance, and independence in terms of geodesic distances and angular metrics. Tools such as the von Mises–Fisher distribution, geodesic means, and angular concentration measures enable coherent inference in these nonlinear domains. Empirical results show that spherical models can match or exceed Euclidean alternatives, particularly in prediction and community detection.

Future work includes adopting variational inference as a scalable alternative to Monte Carlo methods, lowering computational cost while preserving the model’s geometric structure. Another direction is refining MCMC estimation through better tuning of step sizes, proposals, and acceptance criteria, along with adaptive algorithms and parallel chain diagnostics, enhancements that could improve convergence, posterior exploration, and parameter reliability, particularly in low-dimensional or geometrically constrained settings. Methodological extensions include exploring latent structures beyond the symmetric inner product, such as bilinear forms, asymmetric kernels, or other geometric operators, enabling richer relational patterns and the modeling of directed networks by relaxing the symmetry assumption. The framework can also be extended to multilayer and dynamic networks by modeling latent trajectories that evolve over time or span multiple interaction layers, with the geometry adapting to reflect structural changes and contextual information.

From a broader perspective, extending the geometric stock to hyperbolic spaces or other embedded manifolds would allow model geometry to be tailored to specific domains. Such developments would enable systematic study of how geometric assumptions affect fit, predictive accuracy, and interpretability, advancing manifold-based latent space modeling for complex networks.

\section*{Statements and declarations}

The authors declare that they have no known competing financial interests or personal relationships that could have appeared to influence the work reported in this article.

During the preparation of this work the authors used ChatGPT-5 in order to improve language and readability. After using this tool, the authors reviewed and edited the content as needed and take full responsibility for the content of the publication.

\bibliography{references.bib}
\bibliographystyle{apalike}

\appendix

\section{Dirichlet process}

A random distribution function $F$ is said to follow a Dirichlet Process with concentration parameter $\alpha > 0$ and base distribution $G$ on $\mathbb{R}$, denoted $F \sim \textsf{DP}(\alpha, G)$, if for any finite measurable partition $B_1, \ldots, B_k$ of $\mathbb{R}$,  
$(F(B_1), \ldots, F(B_k)) \sim \textsf{Dir}\big(\alpha G(B_1), \ldots, \alpha G(B_k)\big)$. 
Here, $G$ serves as the base (or centering) distribution, while $\alpha$ acts as a precision parameter: larger values of $\alpha$ yield realizations of $F$ that are more tightly concentrated around $G$. See \cite{ferguson1973bayesian} for a detailed discussion of the role of $G$ in the theoretical properties of the DP.

Alternatively, the constructive definition of the DP \citep{sethuraman1994constructive} specifies that $F \sim \textsf{DP}(\alpha, G)$ if, almost surely, $F(\cdot) = \sum_{k=1}^{\infty} \omega_k \, \delta_{\vartheta_k}(\cdot)$,  
where $\delta_{\vartheta}(\cdot)$ is a point mass at $\vartheta$,  
$\vartheta_k \overset{\text{iid}}{\sim} G$, $\omega_k = z_k \prod_{\ell=1}^{k-1} (1 - z_\ell)$, and $z_k \overset{\text{iid}}{\sim} \textsf{Beta}(1, \alpha)$ for $k = 1, 2, \ldots$.  
In this stick-breaking representation, the locations $\vartheta_k$ are drawn i.i.d. from $G$, and the weights $\omega_k$ satisfy $\sum_{k=1}^\infty \omega_k = 1$ almost surely.  
This construction shows that a DP draw is (almost surely) a discrete distribution on $\mathbb{R}$, expressed as a countable mixture of point masses.

\section{Notation}

The cardinality of a set $A$ is denoted by $|A|$. If P is a logical proposition, then $1_{\text{P}} = 1$ if P is true, and $1_{\text{P}} = 0$ if P is false. $[x]$ denotes the floor of $x$. The Gamma function is given by $\Gamma(x)=\int_0^\infty u^{x-1}\,e^{-u}\,\text{d}u$. 
Matrices and vectors with entries consisting of subscripted variables are denoted by a boldfaced version of the letter for that variable. For example, $\mathbf{x} = (x_1,\ldots, x_n)$ denotes an $n\times1$ column vector with entries $x_1,\ldots, x_n$. We use $\mathbf{0}$ and $\mathbf{1}$ to denote the column vector with all entries equal to 0 and 1, respectively, and $\mathbf{I}$ to denote the identity matrix. A subindex in this context refers to the corresponding dimension; for instance, $\mathbf{I}_n$ denotes the $n\times n$ identity matrix. The transpose of a vector $\mathbf{x}$ is denoted by $\mathbf{x}^\top$; analogously for matrices. Moreover, if $\mathbf{X}$ is a square matrix, we use $\textsf{tr}(\mathbf{X})$ to denote its trace and $\mathbf{X}^{-1}$ to denote its inverse. The norm of $\mathbf{x}$, given by $\sqrt{\mathbf{x}^\top\mathbf{x}}$, is denoted by $\|\mathbf{x}\|$.

Now, we present the form of some standard probability distributions:
\begin{itemize}
	
    \item \textbf{Multivariate normal:} A $d\times 1$ random vector $\boldsymbol{X}=(X_1\ldots,X_d)$ has a multivariate Normal distribution with parameters $\boldsymbol{\mu}$ and $\mathbf{\Sigma}$, denoted by $\boldsymbol{X}\mid\boldsymbol{\mu},\mathbf{\Sigma} \sim \textsf{N}(\boldsymbol{\mu},\mathbf{\Sigma})$, if its density function is
    $$
    p(\mathbf{x}\mid\boldsymbol{\mu},\mathbf{\Sigma}) = (2\pi)^{-d/2}\,|\mathbf{\Sigma}|^{-1/2}\,\exp{\left\{-\tfrac12 (\mathbf{x} - \boldsymbol{\mu})^{\textsf{T}}\mathbf{\Sigma}^{-1}(\mathbf{x} - \boldsymbol{\mu}) \right\}}\,.
    $$

    \item \textbf{Beta:}
    A random variable $X$ has a Beta distribution with parameters $\alpha,\beta > 0$, denoted by $X\mid\alpha,\beta\sim\textsf{Beta}(\alpha,\beta)$, if its density function is
    $$
    p(x\mid\alpha,\beta) = \frac{\Gamma(\alpha+\beta)}{\Gamma(\alpha)\,\Gamma(\beta)}\,x^{\alpha-1}\,(1-x)^{\beta-1},\quad 0<x<1\,.
    $$
    
    \item \textbf{Dirichlet:} 
    A $K\times 1$ random vector $\boldsymbol{X} = (X_1,\ldots, X_K)$ has a Dirichlet distribution with parameter vector $\boldsymbol{\alpha} = (\alpha_1,\ldots , \alpha_K)$, where each $\alpha_k > 0$, denoted by $\boldsymbol{X}\mid\boldsymbol{\alpha}\sim\textsf{Dir}(\boldsymbol{\alpha})$, if its density function is
    $$
    p(\mathbf{x}\mid\boldsymbol{\alpha}) =
    \left\{
	\begin{array}{ll}
		\frac{\Gamma\left(\sum_{k=1}^K\alpha_k\right)}{\prod_{k=1}^K\Gamma(\alpha_k)}\prod_{k=1}^K x_k^{\alpha_k-1}, & \hbox{if $\sum_{k=1}^K x_k = 1$;} \\
		0, & \hbox{otherwise.}
	\end{array}
    \right.
    $$
	
    \item \textbf{Categorical:}
    A random variable $X$ has a categorical distribution with parameter vector $\boldsymbol{\pi} = (\pi_1,\ldots , \pi_K)$, where $\sum_{k=1}^K \pi_k =1$, and $\pi_k\geq 0$ for $k=1,\ldots,K$, denoted by $X\mid\boldsymbol{\pi}\sim\textsf{Cat}(\boldsymbol{\pi})$, if its probability mass function is
    $$
    p(x \mid\boldsymbol{\pi}) =
	\left\{
	\begin{array}{ll}
		\prod_{k=1}^K \pi_k^{1_{\{x = k\}}}, & \hbox{if $x\in\{1,\ldots,K\}$;} \\
		0, & \hbox{otherwise.}
	\end{array}
    \right.
    $$
	
\end{itemize}

\end{document}